\documentclass[12pt]{article} 

\usepackage{graphicx}          
\usepackage{multirow}          
\usepackage{amsmath,amssymb,amsfonts} 
\usepackage[title]{appendix}   
\usepackage{xcolor}            
\usepackage{url}               
\usepackage{textcomp}          
\usepackage{manyfoot}          
\usepackage{booktabs}          
\usepackage{algorithm}         
\usepackage{algorithmicx}      
\usepackage{algpseudocode}     
\usepackage{listings}          
\usepackage{mathtools}
\usepackage{geometry}
\usepackage{caption}           
\usepackage{enumerate}         
\usepackage{dirtytalk}         
\usepackage{microtype}         
\usepackage[utf8]{inputenc}    
\usepackage[numbers]{natbib} 
\usepackage{wrapfig}
\usepackage{lmodern}
\usepackage[T1]{fontenc}
\usepackage{subcaption}

\title{Interacting Dark Energy and Its Implications for Unified Dark Sector}

\author{
  Pradosh Keshav MV,  Kenath Arun\thanks{Corresponding author: kenath.arun@christuniversity.in} \\
  \small{\textit{Department of Physics and Electronics}} \\
  \small{\textit{CHRIST (Deemed to be University)}} \\
  \small{\textit{Bangalore, Karnataka, India}}
}
\date{} 

\begin{document}

\maketitle

\begin{abstract}
Alternative dark energy models were proposed to address the limitation of the standard concordance model. Though different phenomenological considerations of such models are widely studied, scenarios where they interact with each other remain unexplored. In this context, we study interacting dark energy scenarios (IDEs), incorporating alternative dark energy models. The three models that are considered in this study are time-varying \(\Lambda\), Generalized Chaplygin Gas (GCG), and K-essence. Each model includes an interaction rate $\Gamma$ to quantify energy density transfer between dark energy and matter. Among them, GCG coupled with an interaction term shows promising agreement with the observed TT power spectrum, particularly for $\ell<70$, when $\Gamma$ falls within a specific range. The K-essence model (\(\Gamma \leq 0.1\)) is more sensitive to \(\Gamma\) due to its non-canonical kinetic term, while GCG (\(\Gamma \geq 1.02\)) and the time-varying \(\Lambda\) (\(\Gamma \leq 0.01\)) models are less sensitive, as they involve different parameterizations. We then derive a general condition when the non-canonical scalar field \(\phi\) (with a kinetic term $X^n$) interacts with GCG. This has not been investigated in general form before. We find that current observational constraints on IDEs suggest a unified scalar field with a balanced regime, where it mimics quintessence behavior at $n<1$ and phantom behavior at $n>1$. We outline a strong need to consider alternative explanations and fewer parameter dependencies while addressing potential interactions in the dark sector.
\end{abstract}
\vspace{0.5cm} 


\newpage

\section{Introduction}
The standard cosmological model, also known as the $\Lambda$-CDM model, has been successful in explaining various observations, ranging from the anisotropies in the cosmic microwave background (CMB) to the large-scale structure of the universe \citep{planck2015cosmological}. However, despite being the most accepted model of the universe, several shortcomings are associated with this model. One particular drawback is the low-$\ell$ anomaly, which arises from the discrepancy between the low amplitude integrated Sachs-Wolfe (ISW) effect predicted by the $\Lambda$-CDM model and the observed data in the lower multipole ranges of CMB data, i.e., $\ell<70$. This inconsistency is attributed to the time-varying gravitational potentials along the line of sight, which can influence the CMB photons as they traverse through large-scale structures in the early universe  \citep{kosowsky2002efficient}. 

To address the low-$\ell$ anomaly, one approach is to introduce a dark energy component that can interact with matter or to modify the theory of gravity \citep{hu2002cosmic}. However, the $\Lambda$-CDM model has other major issues, such as cosmological constant and coincidence problem, further indicating that the current model needs revision \citep{wetterich1988cosmology}.  Recent data from the DESI collaboration hinted at evidence supporting small variations in dark energy over time \citep{abbott2022dark}. This calls for the consideration of dynamical models, such as  Holographic Dark Energy (HDE), New Generalized Chaplygin Gas (NGCG), and the CPL model (also known as the \(w_0w_a\)CDM model), among others. However, such models also have limitations and are sometimes found less effective at fitting data in an economical way \citep{cortes2024interpreting, giare2024interacting}. Consequently, dark energy featuring a time-varying equation of state (EoS) is now seen as necessary. This involves analyzing the EoS of a background fluid as energy density transitions from dark matter to dark energy instead of relying solely on potential-based formulation \citep{sivaram2011dieterici}. 

The Interacting Dark Energy models are primarily designed to elucidate the observed CMB data (particularly at the lower limit) and usually focus on the energy density transfer rate, denoted as $Q$. Based on different types of kernel, several models are proposed \citep{sharov2017new, marcondes2016analytic, feng2016revisit, an2017constraints, caprini2016constraining, jimenez2016cosmological}. For example, Valentino et al. define the kernel $Q$ as the rate at which the energy flow between dark energy and dark matter \citep{diValentino2020nonminimal}\citep{diValentino2017H0}\citep{diValentino2020H0}. This allows $Q$ to take both positive and negative values depending on the direction of flow, i.e., whether the dark energy density is transferred to dark matter $(+Q)$ or vice versa $(-Q)$. This gives a form $ Q = \xi H \rho _x $, where $ \xi $ is a dimensionless coupling constant, $H$ is the Hubble parameter, and $ \rho _x $ is the dark energy density. For different interpretations of IDEs, refer to \citep{wang2016dark}\citep{lucca2020shedding}. Note that the coupling between dark energy and dark matter only modifies the Euler's equation while keeping the continuity equation unaltered \citep{pourtsidou2016reconciling, linton2022momentum, kumar2017observational, jimenez2021velocity}. The coupling of dark energy to radiation \citep{de2018decaying}, the coupling of dark energy to baryons \citep{vagnozzi2020we, beltran2016q}, and the coupling of dark energy to both dark matter and baryons \citep{harko2022observational} are also considered.

The interaction between dark energy and matter components can induce deviations in density evolution and expansion rates from the standard \(\Lambda\)-CDM paradigm. Sometimes, questioning whether fundamental physics even works as usual in the dark sector since, after all, it suffers from the cosmological constant problem  \citep{d2016quantum, farrar2004interacting}. Although promising alternative models like the time-varying cosmological constant, Generalized Chaplygin Gas (GCG), and K-essence could potentially elucidate such dynamics; their applicability is currently limited by observational data \citep{bento2002generalized}\citep{chimento2003link}.  For instance, assuming GCG as an interacting $\Lambda$CDM, a new extension of GCG (known as NGCG) has been widely studied in the literature \citep{barreiro2008wmap, salahedin2020cosmological}. This has a totally different interpretation (a cosmological constant type dark energy interacting with cold dark matter), where the interaction is characterized by a constant parameter. However, it was shown in \citep{dunsby2024unifying, dunsby2024double} that GCG could be considered a limiting case within broader frameworks, such as the Murnaghan EoS or models involving non-minimal coupling between two scalar fields. The non-zero sound speed associated with these generalized models could facilitate early structure formation, which aligns with recent observations \citep{de2024efficient}. This necessitates consideration of a specific EoS resembling a combination of a Chaplygin gas and a non-zero $\Lambda$ that could potentially unify the phenomenological descriptions of different dark energy models into one single fluid. 

Alternatively, one can also formulate the evolution of dark energy EoS based on some principles of quantum gravity, which invoke the holographic dark energy (HDE) model \citep{setare2008phys, setare2009chin, setare2009phys, setare2010jcap, setare2010phys, jamil2009phys, hsu2004phys}. Though HDE based scalar field can be considered equivalent to a modified Chaplygin gas model\citep{pan2012interacting}, its generalizations haven't been explored fully. A correspondence between the tachyon, K-essence, and dilaton scalar field models with the interacting entropy-corrected HDE model in a non-flat FRW universe can be found in \citep{khodam2011commun, karami2011phys}. However, incorporating entropy corrections into the interacting (HDE) model has unpleasant consequences, such as varying universal gravitational constant G, which needs further revision \citep{adabi2012res}. 

In this paper, we introduce a background-dependent parameter ($\Gamma$) to study interacting dark energy scenarios (IDEs) and their constraints. Within an allowed parameter space, $\Gamma$ serves both as a constant and varying interaction rate depending upon the background. Note that we advocate for a positive $\Gamma$ over $Q$ for apparent reasons. For instance, in the GCG model, $\Gamma_{\text{chap}}$ can be interpreted as a parameter that adjusts the interaction strength between dark energy and matter, hence a positive constant. This means that dark energy density in GCG is no longer purely a function of the scale factor $a$ but also depends on $\Gamma$. But while determining the specific value of $ \Gamma_{\text{chap}}$ in scalar dark energy model, such as K-essence, $\Gamma$ is no longer a constant. This is a useful consideration in phenomenologically sound dark energy model \citep{alfano2023dark}, where different parameterizations of $\Gamma$ can be used to probe the early dark sector.

Although GCG and K-essence are popular among cosmologists as a plausible explanation of varying dark energy EoS without invoking a cosmological constant \citep{xu2017cosmological}, we aim to go further by tracking the rate of interaction, the entropy of dark energy and matter and their densities over conformal time. The potential possibility of IDEs suppressing the matter power spectrum has already been studied in the literature \citep{farrar2004interacting, buen2018interacting, xia2009constraint, barkana2018possible, caldera2009growth, lesgourgues2016evidence}. Therefore, a detailed analysis would require studying the evolution of dark energy density in both linear and nonlinear regimes, opening the possibility of unifying alternative dark energy models emerging from a single scalar.

The paper is organized as follows. We start by examining the early background of the universe in section 2. We then study the dynamics of an interacting FLRW background using a lapse function. Afterward, we track the entropic density of matter, radiation, and dark energy to establish a relationship between the interaction rate for a specific thermodynamics case, which is discussed in sections 3 and 4. In section 5, we derive an expression for the interaction rate and examine the dark energy density in relation to the background fluid density, with a varying EoS that is sensitive to the interaction rate. This approach is further applied in section 6 to discuss the behavior of GCG and K-essence at early times. However, section 7 introduces a different K-essence Lagrangian that reduces to the same dark energy EoS at late times. We analyze theoretical predictions with observational data from SNe and CMB in section 8, followed by a detailed discussion in the last section. Since the primary focus of this study will be on the sensitivity of $\Gamma$ and its implications to unified dark energy models, we will not be discussing structure formation and cold dark matter scenarios in this context.

\section{The FLRW Background and Early Universe}

The measurements \citep{perlmutter1999measurements}\citep{riess1998observational} from Type Ia supernovae famously pointed out that the universe is expanding at an accelerated rate. By measuring the flux of galaxies with redshift $z$, the deceleration parameter, $q_0$, was found to be $-0.51$.  A negative $q_0$ suggests an accelerated expansion popularly attributed to dark energy. To explain the negative deceleration parameter, one way is to add a repulsive gravity source coming from the EoS of the background. \citep{jones2004introduction}.

The deceleration parameter can be written as:
\begin{equation}
q = \frac{1}{2} (1+3w)\left[1+ \frac{k}{(aH_0)^2}\right]
\end{equation}
where $a$ is the scale factor, $w=w_m + w_r + w_{de}$ is the EoS of the background space-time, $H_0$ is the Hubble constant, and $k$ is the curvature parameter, which takes values 1, 0, and -1. From the background FLRW space-time\citep{Weinberg1973GravitationAC},
\begin{equation}
ds^2 = -N(t)dt^2 + a^2 (t)dx^2
\end{equation}
The critical density ($\rho_c = \rho_{total}$) arising from this background indicates a flat universe. Observations of anisotropies in the CMB analyzed by the Planck team confirm this claim \citep{aghanim2020planck,aghanim2020planckv, debernardis2000flat}. However, supporting this in favor of the $\Lambda$CDM model is challengin since the discrepancy in the model is high for multipole values $\ell<70$ as it requires an understanding of the coincidence problem as well.

The interaction between dark energy and matter with a non-minimal coupling brings an interesting explanation to this scenario (refer to \citep{chimento2003interacting}\citep{wang2005transition}\citep{wei2007observational}\citep{kumar2016probing}\citep{amendola2007consequences, das2006superacceleration}). It follows that dark energy and matter are not merely interacting through gravity but can also be within the scope of a hypothetical fifth force or interaction. Especially when the appropriate interaction model can produce an effective dark energy EoS, similar to the quintessence model \citep{wetterich2004phenomenological}. We aim to investigate this scenario further without assuming any fifth force or a universal interaction term but by looking for background-dependent parameters non-minimally coupled to any scalar field.

We take into consideration the lapse $N$ function and a time-dependent scale factor $a(t)$ at equilibrium, with a simple variation that accounts for early dark energy:
\begin{equation}
\frac{\delta}{\delta N} \xrightarrow{} \frac{3}{N^2} m_{Pl}^2 a \Dot{a}^2 = a^3 \rho(a)
\end{equation}
The Friedman equation gives the relationships between the Hubble constant $H_0$ and the total matter-energy density, including dark energy. We can modify this equation to include the Planck mass $m_{Pl}$, where $\rho_{}$ represents the background density per unit volume:
\begin{equation}
\frac{\Dot{a}^2}{N^2 a^2} = \frac{1}{3 m_{Pl}^2} \rho_{} = \frac{8 \pi G}{3} \rho_{total} - \frac{k}{a^2}
\end{equation}

Usually, we set the value of $N=1$ \citep{MELIA2019167997} and combine the Hubble parameter $H$ with the variation of a lapse function. The conservation equation becomes:
\begin{equation}
\Dot{\rho} = - 3H(\rho + p)
\end{equation}where we equate $p= w \rho$. This yields:
\begin{equation}
\Dot{\rho} = -3H\rho (1+w)
\end{equation}
which is generally considered as the density evolution of the background metric.

From a good relativistic approximation\citep{ceylan2015relativistic}, the observational constraints from $\Lambda$-CDM usually take different values for the background: for matter-dominated era, $w_m = 0$ and $\rho_m \propto (1+z)^3$; for radiation-dominated era, $w_r = 1/3$ and $\rho_r \propto (1+z)^4$; for dark energy-dominated era, $w_{de} = -1$ \citep{arun2017dark}. The notations follow standard form where the subscript ({de, m, r}) represents dark energy, matter, and radiation components. In the coming sections, we will discuss scenarios where the constrained deceleration parameter varies with different eras.

\section{Dynamics of Interacting FLRW Background}
Interacting dark energy models offers a fresh perspective to cosmology as well as high energy physics \citep{Sola2015, Begue2019}. These models often conflict with boundary horizons because they require a coupling term that balances the ratio of dark energy to dark matter at late times \citep{hu2006interacting}. The assumption of a primordial plasma (under thermal equilibrium) between dark matter and baryonic matter simplifies the equation of motion by neglecting the boundary terms \citep{pietroni2009non}. The equation of motion for the background space-time can be derived from the Einstein-Hilbert action, which describes the dynamics using the lapse function and a flat spatial metric (k=0). In this scenario, one can calculate the number of interactions to derive the deceleration parameter based on the entropic densities of matter, radiation, and dark energy \citep{baer2015dark}.  

For a flat spatial metric, we usually take the lapse function of the signature metric $(+,-,-,-,-)$ given by the Einstein-Hilbert action \citep{weinberg2008cosmology}. Here, we further ignore the contribution of boundary terms to arrive at the moduli space approximation. In a particular case where the primordial plasma is initially stable, the equilibrium condition for a matter-dominated universe is obtained by setting $\rho= \rho_r + \rho_m + \rho_{de}$ in the action functional $\mathbf{S}$.

This allows us to write the action as:
\begin{equation}
\mathbf{S} = \int d^4 x \left(\frac{m_{pl}^2}{2}(\frac{-6a\Dot{a}^2}{N})-a^3N(- \rho(a))\right)
\end{equation}The entire dynamics would be different if the plasma got destabilized in a radiation-dominated era, i.e., $a(t) \propto t^{1/2}$. Similarly, we have $a(t) \propto t^{2/3}$ for the matter-dominated era and $a(t) \propto \exp(H(t))$ for the dark energy-dominated era. 

The Friedman equation can be modified for large-scale structures based on different components of dark energy, matter, and radiation. By approximating the total energy density of each component to zero, equation (4) can compare the characteristics of matter and radiation with that of dark energy in different epochs of the universe. 

The density parameters for different epochs are:
\begin{equation}
\Omega_{i}(z) = \frac{\rho_r + \rho_m + \rho_{de}}{\rho_{\text{critical}}}
\end{equation}
where $\Omega_{i}(z) = 1$ shows a flat universe. From equation (4), we substitute $\rho(z) = \rho_{m}$ to yield the equilibrium condition with critical density:
\begin{equation}
\frac{8 \pi G}{3} \rho_{}(z) = \frac{1}{3 m_{Pl}^2} \rho_{\text{total}}
\end{equation}
By rearranging the terms, we get the critical density for matter as:
\begin{equation}
\Omega_m (z) = \frac{\rho_m (z)}{\rho_{\text{critical}}} = \frac{H_0^2}{ 8 \pi G m_{Pl}^2} \approx 0.271 \pm 0.044
\end{equation} which gives us the value of $\Omega_m (z) = 0.315 \pm 0.007$ at present epoch (i.e., $z=0$)\citep{aghanim2020planck}. 

By substituting equation (10) into (1), we get the deceleration parameter:
\begin{equation}
q(z) = \frac{-\ddot{a}}{aH_{0}^2} = \frac{1}{2} \sum_i \Omega_{i}(z) [1+3w_i (z)]
\end{equation} where we get the constrained values:
\begin{equation}
    \frac{q_m(z)}{q_i(z)}= 3, \quad \frac{q_r(z)}{q_i(z)}=4, \quad \frac{q_{de}(z)}{q_i(z)}=-6
\end{equation}We use these values to model the interaction dynamics of the early universe. 

During the radiation and matter-dominated eras, gravity could slow down the expansion. This implies:
\begin{equation}
    \frac{q_m(z)}{q_i(z)} >0, \quad \frac{q_r(z)}{q_i(z)}>0
\end{equation}We take $\frac{\ddot{a}}{a}=-3H_{}^2$ and $\frac{\ddot{a}}{a}=-4H_{}^2$ (for a non-accelerating type with $\ddot{a}<0$), respectively. Thus, in the dark energy-dominated era:
\begin{equation}
    \frac{q_{de}(z)}{q_i(z)}<0, \quad \ddot{a}>0
\end{equation}we have an accelerating phase with $\frac{\ddot{a}}{a}=6H_{}^2$. Here, the double derivative of the scale factor with respect to time gives the Raychaudhuri equation:
\begin{equation}
\frac{\Ddot{a}}{a} = -\frac{4 \pi G}{3} (\rho + 3P)
\end{equation}which is the rate of change in the expansion or contraction of the universe. Note that the current observational data for this value are not exact and typically vary for different cosmological models.

For a dark energy-dominated era, one can find $q_{de}$ by solving equation (15) for the current value of $w_i(z)$. Given that we assume a flat universe with only matter (including dark matter) and dark energy, the best estimates of the Hubble constant is around $H_0= 73.5 \pm 0.9$ km/s/Mpc and $q_{de}$ is approximately $-0.51 \pm 0.024$ \citep{riess2022comprehensive}. It may change by adding other components, such as radiation and curvature.

\section{Dark Energy and Matter Interaction in the Early Universe}
Interacting dark energy models were initially proposed to explain the small value of the cosmological constant (refer to \citep{gavela2009dark, gavela2010dark, di2020interacting, di2017can, di2020nonminimal, lucca2021multi} for more literature). Subsequently, it became useful to examine varying EoS in a dynamic background, thus offering a natural solution to the coincidence problem \citep{yang2018interacting}. However, in our context, two major approaches can be used to study the interaction between matter and dark energy in the background space-time. The first approach considers the background space-time as elastic, in which dark energy and dark matter are modeled as disturbances in this elastic system, and their effect on the evolution of the universe is determined by the properties of this background \citep{linton2022momentum,kumar2017observational}. The second approach considers the scattering of dark matter particles with standard model particles \citep{doglioni2016dark}. 

We assume that dark energy can be measured by its interaction with matter for a given rate of interaction $\Gamma$ and subsequent change in the EoS of background fluid. This interpretation of $\Gamma$ differs from the interaction kernel $Q$, where the former specifically represents the interaction rate specified by the dark energy model under consideration. Here we specifically choose a positive $\Gamma$ with a thermodynamic cut-off for the dark energy component expressed in terms of the Hubble scale \citep{cruz2018holographic}. 

The number of interactions $N'$ is given by:
\begin{equation}
N' = \int_{t_0}^{t} \Gamma dt = \int_{T_1}^{T_2} \frac{dt}{dT}dT
= \int_{T_1}^{T_2} \Gamma \frac{T}{\Dot{T}} \frac{dT}{T}
\end{equation}
where $T$ is the temperature of the primordial plasma and $\Gamma>0$ determines the rate at which energy is transferred from dark energy to matter. Assuming that the matter and radiation were at equilibrium with dark energy at an early time $t_0 <t$, we can infer $T_1 = 2T_2$, i.e., the universe is doubled by a factor of 2. This simplifies equation (13) to:
\begin{equation}
N' = - \int_{T_1}^{T_2} \Gamma H^{-1} \frac{dT}{T}
\end{equation}
Here, we can take $N' = \frac{\Gamma}{H} \ln(2)$ for $\Gamma > H$ at thermal equilibrium and $\Gamma < H$ for all non-equilibrium conditions. 

By substituting the value of the scale factor $a$ into the above equation, we get:
\begin{equation}
N' = \Gamma (\frac{\Dot{a}}{Na})^{-1}
\end{equation}
and
\begin{equation}
N' = \Gamma H^{-1}
\end{equation}
It is interesting that the number of interactions $N'$ in equation (19) is not directly related to the action functional in equation (7) but through the Hamiltonian constraint in the FLRW interacting background. Thus, one can relate the Hamiltonian density to the Lagrangian density by modifying the action functional.

On the other hand, $t >>t_0$ is related to the entropy of the early universe. With an increase in entropy for positive timescales, fluctuations in matter entropic density of the primordial plasma become pivotal for structure formation \citep{frautschi1982entropy}. We typically consider the conformal time $Ndt = d\tau$ when the universe is eight-fold larger.

Given some physical degrees of freedom, one can prescribe that macroscopic physical properties like thermodynamic properties could emerge from an interacting background \citep{capozziello2018information}. This means the entropic density follows $S \equiv \frac{\rho + P}{T} $.
By analyzing the entropy density of matter, radiation, and dark energy at $T_1 = 2T_2$, where the size of the plasma sphere is doubled to its initial size. We can compute the entropic densities for different epochs as:
\begin{align}
    S_{m} &\approx \Gamma^{ \frac{q_m}{q_i}} N dt \Rightarrow S_{m} \approx \Gamma^{ 3}  d\tau \notag \\
    S_{r} &\approx \Gamma^{ \frac{q_r}{q_i}} N dt \Rightarrow S_{r} \approx \Gamma^{ 4}  d\tau \\
    S_{de} &\approx \Gamma^{ \frac{q_{de}}{q_i}} N dt \Rightarrow S_{de} \approx \Gamma^{-6}  d\tau \notag
\end{align}Here, we consider an epoch where $S_m >S_r + S_{de}$ and $q_m < q_i$, which is significant for a self-interacting dark sector that possibly avoids complexities due to quantum decoherence \citep{kiefer2011cosmological, capozziello2013dark}. For example, a popular dark matter candidate known as ultralight axions is often modeled with a lower $q_m(z)$. Decoherence occurs through the gravitational interaction of the axion overdensity with its environment, which is provided by baryonic matter \citep{allali2020gravitational}. This helps to constrain axion-like particles with mass parameter range $10^{-33}eV \leq m_a \leq 10^{-18}$ eV with lower bound typically in the range of $H_0/h \sim 100 $ Km/s/Mpc constrains axion dark energy at early times \citep{marsh2016axion}. 

The Hubble parameter can be modified as a power law in the form: \begin{equation}
    H^2  d \tau^2 = \Gamma^{( \frac{1}{2} \Sigma_i \Omega_{m}(z) [1+3w_i (z)] )} 
\end{equation} which serves as a constrained parameter in a matter-dominated universe for all $N \neq 1$. For the early universe, where matter-radiation densities are dominated, the total density relation with redshift can be derived by substituting $S_m$ and $S_r$ from equation (20) to equation (11):
\begin{equation}
    \rho_i (z) \approx \frac{3 m_{pl}^2}{d\tau^2 }    \Gamma^{( \frac{1}{2} \Sigma_i \Omega_{m}(z) [1+3w_i (z)] )}
\end{equation} The constraint for $q_i (z)$ represents a linear density perturbation term ($\rho_i$) for a given component $i$ at redshift $z$, at which the scale factor of the universe evolves as a function of time. Since we already constrained values from equation (17) that allow the energy density of matter scales as $\rho_m (z) \propto \Gamma^3$, we can also compute contributions from other major components. The energy density of radiation scales as $\rho_r (z) \propto \Gamma^4$, and the energy density of dark energy scales as $\rho_{de} (z) \propto \Gamma^{-6}$.

We note that these constrained values imply a background consisting of multiple fluids with varying EoS ($w_i$). The density parameter ($\rho_i$) should be calculated as the product of a parameter ($\Gamma$) that scales to the net sum of matter density and EoS parameters. From our best estimates, we can approximate the value of $w$ to be approximately equal to -1. 

For an accelerating universe, we can consider $\frac{\ddot{a}}{a} = 6H^2$ and $\Omega_{m}(z)=0.3$. This simplifies equation (22) to:
\begin{equation}
    \rho_i (z) \propto \frac{3m_{pl}}{d\tau^2 } \ \Gamma^{-0.3}
\end{equation}
which is the total density of the background with respect to redshift $z$ and the rate of interaction between matter and dark energy. Moreover, $rho_i(z)$ decreases with the rate of change of scale in $\Gamma$, as it is a good candidate to solve the coincidence problem as well \citep{wang2014cosmological}.

It is worth mentioning that during the earlier stages of the universe, the impact of dark energy on equation (23) is minimal, which is significant for the formation of structures. This is mainly because the exponent of $\Gamma$ is linked to the rate of total interactions and only scales with the contribution of matter density during the early times of the universe. The changing scales of interaction can influence the evolution of the total matter density term of the background fluid (and not dark energy) over a positive timescale.
\section{Dark Energy, Entropy, and the Dynamics of Background}
\subsection{Dark Energy, Entropy, and Expansion Rate}
The holographic principle, which posits that the entropy of a system scales with the area of its boundary rather than its volume, has been invoked in various aspects of cosmology, particularly for black hole entropy \citep{susskind1995world}. It has been suggested in \citep{sivaram2013holography} that the presence of large-scale dark energy, which is restricted by the hierarchy of structures, might imply a similar notion of entropy for these structures proportional to their surface area. While black hole entropy is well-established, applying this concept to cosmic structures distinct from black holes requires further motivation and development. Primordial structures might follow a similar entropy-area relation \citep{banks2001holographic, banks2017holographic, nojiri2022early, mcfadden2010holographic, mcfadden2010holography}. This necessitates considering the interaction between the primordial background and dark energy, particularly when considering the gravitational self-energy density of these structures must at least equal or exceed the repulsive dark energy density, commonly denoted by \(\Lambda\) \citep{sivaram2008scaling, sivaram2012minimum}.

For cosmic structures, the entropy $S$ is given as:
\begin{equation}
    S= K_B \frac{{\Lambda}}{\left(G m / c^2\right)} (4 \pi R(i)^2)
\end{equation} where $R(i)(=ct)$ is the radius of range of interaction of dark energy component (bound to Hubble horizon), $k_B$ is Boltzmann’s constant, $c$ is the speed of light, $G$  is the gravitational constant, $m$ is the mass of the particle (both baryonic and dark matter). While black hole entropy can be conceptualized as the total number of \say{Planck areas} that cover the horizon's surface \citep{bekenstein1973black, bekenstein1975statistical, sivaram2009curvature}, the entropy for cosmic structures constrained by dark energy can be determined by the range of interaction covering the surface area of primordial plasma. Primarily because the area that scales to the range of interaction between dark energy and matter can be taken as $\ \frac{{\Lambda}}{\left(G m / c^2\right)}$, which is a constant. Hence, it is reasonable to hold $\Lambda$ accountable for the primordial curvature that approaches zero at the horizon boundary when $R^2$ is maximum \citep{sivaram1994fundamental}, indicating a constant dark energy density. This can be written as:
\begin{equation}
S_{de} =   \frac{3k_B c^2}{G m} (4 \pi R^2(i)) \rho_{de}
\end{equation}
We can substitute $m=m_{Pl}$ in equation (25) with the entropic density of de from equation (20) to obtain:
\begin{equation}
\Gamma^{-6} d\tau = \frac{3k_Bc^2}{G m_{pl}} (4 \pi R^2(i)) \rho_{de}
\end{equation}
where $d\tau$ is the conformal time. The above equation can be rearranged to get the dark energy density of the background as:
\begin{equation}
{\rho_{de}} = \frac{G m_{pl}}{3 k_B c^2} \left(\frac{1}{4 \pi R^2(i) \Gamma^6 d\tau}\right)
\end{equation}

The entropy related to dark energy that depends on the cross-sectional area in a way that $S_{de} \simeq \Gamma^{-6} d\tau \propto R^2$, where the rate of interaction of dark energy per unit volume is proportional to the surface area. The derivative of $\rho_{de}$ w.r.t. $d\tau$ gives:
\begin{equation}
\frac{\dot{\rho_{de}}}{\rho_{de}} = -2 \frac{\dot{R}}{R} - \frac{2}{3} \ \frac{d \ (ln \ \Gamma)}{d \ (ln \ a)} \ \frac{\dot{a}}{a}
\end{equation}The above relation follows equation (20) and represents the rate of change of dark energy density (w.r.t. current dark energy density $\rho_{de}$) in terms of the scale of interaction and the expansion rate of the Universe.

\subsection{The Lagrangian Density and Varying EoS}
The Lagrangian density for a perfect fluid can be expressed as $\mathcal{L} = -\rho c^2 - p$, where $\rho$ denotes the energy density and $p$ is the pressure of the fluid. The EoS parameter, denoted by $w$, relates the pressure of dark energy, $p_{de}$, to its density, $\rho_{de}$, through $p_{de} = w \rho_{de} c^2$. Since dark energy is both homogeneous and isotropic, it is position-independent and has zero viscosity; thus, $w$ remains nearly constant.

By substituting equation (27) for $\rho_{de}$, we get:
\begin{equation}
    p_{de} = w \rho_{de} c^2 = w \left(\frac{3 k_B c^4} { G m_{pl}} R^2(i) \Gamma^6 d\tau \right)^{-1}
\end{equation}
and we can find the Lagrangian density for the given dark energy EoS as:
\begin{equation}
    \mathcal{L}_{de} = -\left(1 + w\right)\left(\frac{3 k_B c^4} { G m_{pl}} R^2(i) \Gamma^6 d\tau \right)^{-1}
\end{equation}
Here $\mathcal{L}_{de}$ is the Lagrangian for the dark energy component from which we derive the action for the same: 
\begin{equation}
    \mathbf{S}_{de} =  \int  \mathcal{L}_{de} dt
\end{equation} \newline
To vary this action with respect to $\Gamma$, we express $ \mathbf{S}_{de}$ in terms of varying $\Gamma$ by using the relation $N \propto 1/\Gamma$. The integral of fourth-order differentiation equation is given by:
\begin{equation}
    \mathbf{S}_{de} = -\left(1 + w\right)\frac{G m_{pl}} { 3 k_B c^4} \int d^4x \sqrt{-g} \Gamma^{-7} R^{-2}(i)
\end{equation}Here, we can vary the action for $\Gamma$ as the interaction changes over time in the presence of matter and radiation fields. This variation gives the equation of motion: \begin{equation}
  \Gamma = \left[\frac{(1+w)}{R^2(i)}\frac{7G m_{pl}}{3 k_B c^4} \sqrt{-g}\right]^{1/8}
\end{equation}If the parameter $\Gamma$ is dynamic, the background metric changes over time in accordance with the interaction scale described by $R^2(i)$. However, in a dark energy model where $\Gamma$ is not dynamic, it does not affect the equation of motion as the action does not explicitly depend on the lapse. If the EoS varies, we have varying $\Gamma$ in the following form:
\begin{equation}
    \frac{\dot{\Gamma}}{\Gamma} = \frac{1}{8} \ \frac{\dot{(1+w)}}{(1+w)}-\frac{1}{4} \ \frac{\dot{R}}{R}
\end{equation}
\begin{figure}[t]
    \centering
    \begin{subfigure}{0.5\textwidth}
        \centering
        \includegraphics[width=\textwidth]{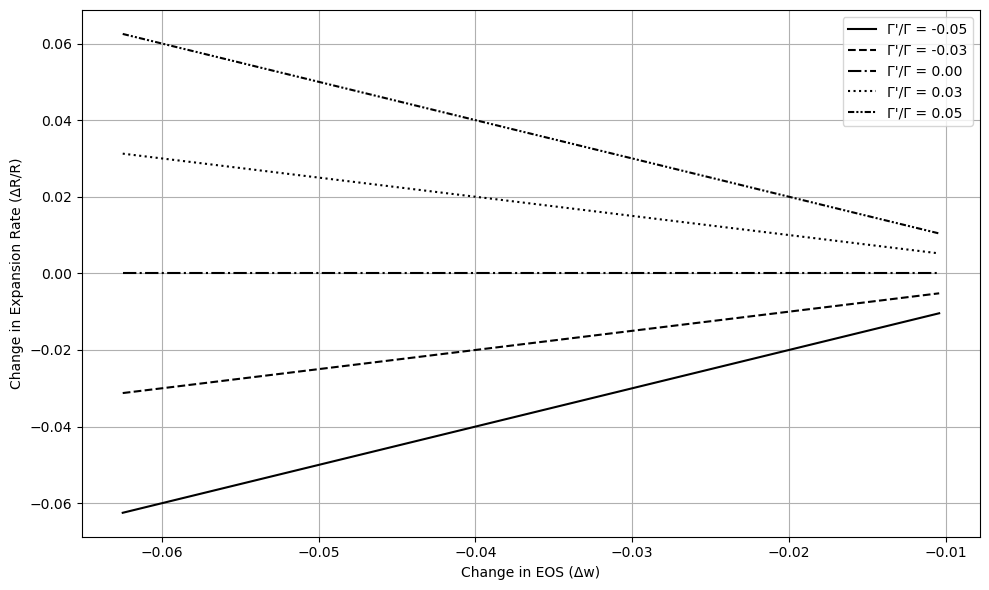} 
        \caption{\small Weak constraints on interaction rate}
    \end{subfigure}
    \hspace{2mm}
    \begin{subfigure}{0.5\textwidth}
        \centering
        \includegraphics[width=\textwidth]{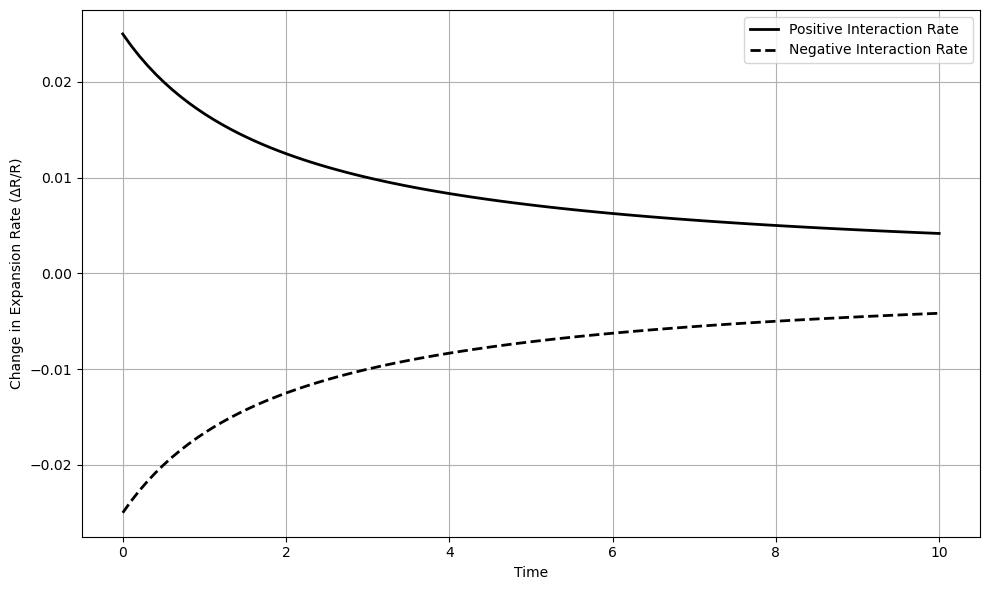} 
        \caption{\small Effect of interaction rate on expansion}
       
    \end{subfigure}
    \caption{\small Illustrates the sensitivity of $\Gamma$ in relation to change in EoS $\Delta w$: Figure (a) shows the variation in the EoS (unstable when $w_0 \sim -1$ \citep{yang2018interacting}) in a flat FLRW background with an interaction rate constrained for different ranges of $\Delta R$; converge close to $\Delta w = -0.1$. The change in $\Gamma$ ranges from $-0.05$ to $0.05$, with an initial value of $w_0 = -0.8$. Figure (b) shows a behavior similar to an inflationary universe where the interaction rate $\Gamma$ steadily increases in response to the change in the expansion rate.}
    \label{fig:sensitive gamma}
\end{figure}
The equation above offers insight into the fact that the range of values for $\Gamma$ cannot be well constrained, mostly because it depends on two other weak constraints. As an alternative approach, we propose IDEs in the upcoming sections to constrain $\Gamma$ for a specific dark energy EoS.

There is a neutralizing adjustment in the interaction rate that is very close to zero, consistent with the currently accepted value of $w$ ( $\approx -1$ \citep{aghanim2020planckv}). Since the EoS parameter is directly proportional to the interaction rate, even a small change in $\Delta w$ can significantly impact the expansion rate, as shown in Fig. \ref{fig:sensitive gamma}(a). For example, if the value of $w$ changes from $-0.8$ to $-0.9$, the interaction rate will change from negative to positive (see Fig. \ref{fig:sensitive gamma}(b)). This change in the interaction rate can either accelerate or decelerate the metric, depending on whether the interaction rate is positive or negative.
\section{Interacting Dark Energy Models}
The $\Lambda$-CDM model faces challenges such as the cosmological constant, coincidence, and fine-tuning problems. These issues have led to the study of alternative models such as time-varying cosmological constant, Generalized Chaplygin gas, and K-essence models, among others \citep{arun2017dark}. Most of these models, however, do not account for interacting dark energy and dark matter cases. The interaction between dark energy and matter is central, as it allows for the study of the evolution and structure formation in the early universe by imposing constraints on model parameters \citep{wang2016probing} \citep{costa2017probing}. Moreover, its absence can lead to inconsistencies when comparing the models to observational data. In an attempt to address these shortcomings, we introduce an interaction term into some of the alternative models. We expect it to potentially resolve certain issues related to original models, such as stability, setting a specific dark energy EoS, and so on  \citep{carneiro2014dark}, as discussed in the previous section. 

\subsection{Time-Varying Cosmological Constant}
A time-varying cosmological constant has been previously introduced in the literature \citep{sivaram2020a} to explain the challenges arising from the initial phase of the Universe, where a Planck scale inflation was proposed. Similar arguments \citep{rebecca2021time} also lead to constraints on the sizes of large-scale structures, such as galaxies and clusters at high redshifts, which are closer to their actual observed sizes. Building on this, we examine the impact of a time-dependent cosmological constant on the expansion rate of the Universe. The model features an almost constant equation of state (EoS) of $w=-1$, which is equivalent to a varying cosmological constant.

Substituting the expression for $\rho_{de}$ into:\begin{equation}
\rho = \rho_{m} + \rho_{de} = \rho_{m} +  \ \kappa \left( \frac{1}{ R^2(i) \Gamma_{\Lambda(t)}^6 d\tau}\right)
\end{equation} where $\kappa = \frac{ G m_{Pl}}{3 k_B c^2}$ which we use as a notation from hereafter. We can substitute $d\tau = Ndt = t_0$ with lapse $N=1$ into equation (35) to get the first Friedmann equation:
\begin{equation}
    H_{}^2= \frac{8 \pi G}{3} \rho_{m} + \frac{8 \pi G}{3} \  \kappa \left( \frac{1}{ R^2(i) \Gamma_{\Lambda(t)}^6 d\tau}\right) - \frac{kc^2}{a^2}
\end{equation} 

If we assume a small value of $\Gamma_{\Lambda(t)} = 0.01$ (which can be later inferred from Figure \ref{fig:baysien}), we yield:
\begin{equation}
    \Lambda(t) \simeq \frac{8 \pi G}{3} \  \kappa \left( \frac{1}{ R^2(i) \Gamma_{\Lambda(t)}^6 d\tau}\right)  \approx 10^{-52} \ m^{-2}
\end{equation}which represents a time-dependent cosmological constant $\Lambda(t) = \Lambda_0 \Gamma(t)^6$ in terms of other fundamental constants and parameters in the FRW background metric. Also, $\Lambda$ changing with $R^{-2}$ can change the predictions of the standard cosmology in the matter-dominated epoch \citep{chen1990implications}. The Hubble parameter becomes:
\begin{equation}
      H_{}^2= \frac{8 \pi G}{3} \rho_{m} +  \frac{\Lambda (t)}{3} \Gamma_{\Lambda(t)}^6 d\tau - \frac{kc^2}{a^2}\Gamma_{\Lambda(t)}^6 d\tau
\end{equation}

It is important to note that in a $\Lambda(t)$ model, we assume that matter and dark energy interact minimally. Where we can normalize $\Gamma_{\Lambda(t)}^6 d\tau$ by integrating it over the ranges of 0 and $\tau_0$, assuming its contribution is negligible during the matter-dominated phase. The power law form consistent with coupled field equations will become:
\begin{equation}
    \Gamma(t) =\Gamma_0 (t/t_0)^6
\end{equation}
By using the relation $a(t) = (t/t_0)^n$, where $t_0$ is the Universe's current age, and $n$ is a constant, we start at some early time $t_{min}$, to calculate this integral as follows:
\begin{equation}
\begin{split}
\int_0^{\tau_0} \Gamma^6 \, d\tau &= \int_0^{\tau_0} \left(\frac{\dot{a}}{a}\right)^6 \, dt \\
&\Rightarrow \int_0^{\tau_0} \left(\frac{n}{t}\right)^6 \, dt \\
&= n^6 \int_0^{t_0} t^{-6} \, dt \\
&\Rightarrow \left[ -\frac{n^6}{5 t^5} \right]_{t_{min}}^{t_0} \\
&= \frac{n^6}{5} \left( \frac{1}{t_{min}^5} - \frac{1}{t_0^5} \right)
\end{split}
\end{equation}

 For normalization, we can set:
\begin{equation}
    \frac{n^6}{5}\left(\frac{1}{t_{min}^5 - t_{0}^5} \right) = 1
\end{equation}
However, such a normalization does not change the value of the theoretical energy density, which can be calculated from equation (38). Additionally, normalization does not accurately capture the physical evolution. Instead, this model can be used to study the behavior of $ \Gamma $ in coupled field equations. To accurately represent the physical evolution, it is essential to solve the complete set of coupled equations.

Suppose $n$ represents the exponent of the scale factor in the Friedmann equation. We can analyze how $\Gamma$ influences the expansion rate and the evolution of the scale factor by solving it for different integer values of $n$. When we substitute $a(t) = (t/t_0)^n$, we obtain:
\begin{equation}
H^2 = \frac{8 \pi G}{3} \rho_{m} + \frac{1}{3} \Lambda_0\left(\frac{ t_0^{2n}}{t^{2n}}\right)  
\end{equation}
Solving the above equation for different integer values of $n$ would give us the corresponding values of $H$ and the evolution of the scale factor of the Universe. 

For $0 < n < 2/3$, the universe enters an accelerating phase. Around $n \sim 2/3$, the scale factor follows a power law consistent with the early Universe's inflationary model. When $n \sim 0$, the scale factor changes linearly with time, representing a static universe. In cases where $n$ is a positive value, the universe's energy density decreases more rapidly than the inverse of the scale factor, a characteristic found in many dark energy models like quintessence \citep{frieman2008dark}.  
\subsection{Generalized Chaplygin gas}
The Generalized Chaplygin gas model \citep{chaplygin1904} is a specific example of a larger class of models with the property of unifying dark energy and dark matter. Following Chaplygin, Kremer proposed that the Chaplygin gas equation of state could be a candidate for dark energy interacting with dark matter and neutrinos in a uniform, flat, isotropic Universe \citep{kremer2007dark}. There are other models with similar features, such as the Tachyon scalar field model and the Dilaton model \citep{kamenshchik2001alternative}. 

The traditional Generalized Chaplygin Gas (GCG) model is characterized by the equation of state:
\begin{equation*}
p_{chap} = -\frac{A}{\rho_{\text{chap}}^{\alpha}}
\end{equation*}
where $A$ is a positive constant, $\alpha$ is a free parameter, and $p_{\text{chap}}$ and $\rho_{\text{chap}}$ are the pressure and energy density of the Chaplygin gas, respectively. To simplify the model, we introduce an interaction rate \(\Gamma_{\text{chap}}\), modifying the equation of state for the dark energy component:
\begin{equation}
p_{de} = -\frac{A}{\rho_{de}\Gamma_{\text{chap}}^{6}}
\end{equation}
and $\Gamma_{\text{chap}}$ represents an effective interaction rate combining both matter (baryonic and dark matter) and neutrino interactions. Note that this won't always be the case, as we consider it for simplicity. 

The modified equation (43) resembles the van der Waals equation, where \(\Gamma_{\text{chap}}\) can be seen as influencing the pressure and volume terms, analogous to how intermolecular forces and finite molecular sizes modify the ideal gas law. In this context, \(\Gamma_{\text{chap}}\) could be interpreted as a parameter that adjusts the interaction strength between dark energy and other components of the universe.

Moreover, \(\Gamma_{\text{chap}}\) encapsulates the interaction dynamics that \(\alpha\) would traditionally describe, reducing the number of free parameters. By comparing the above equations, we can derive an expression for $\rho_{de}$ as a function of $\Gamma$:
\begin{equation}
\rho_{de} = \sqrt{A+ \frac{B}{a^6}}
\end{equation}
With the value of the Chaplygin gas constant,
$$A =\frac{1}{\kappa}R^2\Gamma_{\text{chap}}^6 d\tau$$ we get a van der Waals type of gas:
\begin{equation}
\rho_{de} = \sqrt{ \frac{1}{\kappa}R^2\Gamma_{\text{chap}}^6 d\tau + \frac{B}{a^{6}}}
\end{equation}
Here, the varying $A$ represents the contribution from the interacting component of the dark energy, which is assumed to have an energy density proportional to $\Gamma^{6}$. $B$ is an integration constant ($\sim$ 0.06 \citep{thakur2009modified}) with a positive value representing the contribution from the non-interacting dark energy component with a constant energy density. 

For small values of scale factor $a$, we can deduce that the non-interacting part of dark energy is dominant, and the interacting part only becomes significant later on. When the scale factor $a$ is large, we enter a phase where dark energy dominates, and we have $\rho = \rho_{de} \sim \sqrt{A} \Gamma^{-6}$, similar to time-varying cosmological constant. It's worth noting that $\rho \sim \sqrt{A}$ satisfies the equation $\rho+ p = \rho - \frac{A}{\rho} = 0$. This can be understood as a quintessence model for the equation of state \citep{bento2002generalized}, wherein the interacting component of dark energy becomes dominant over the non-interacting component at later times.

Combining the matter and dark energy density components yields FRW Chaplygin background that retains its energy density as a function of the scale factor:
\begin{equation}
H_{}^2= \frac{8 \pi G}{3} \left( \rho_{m} + \sqrt{A + \frac{B}{a^{6}} }\right)
\end{equation}
\begin{figure}
  \centering
  \begin{subfigure}[b]{\textwidth}
    \includegraphics[width=\textwidth]{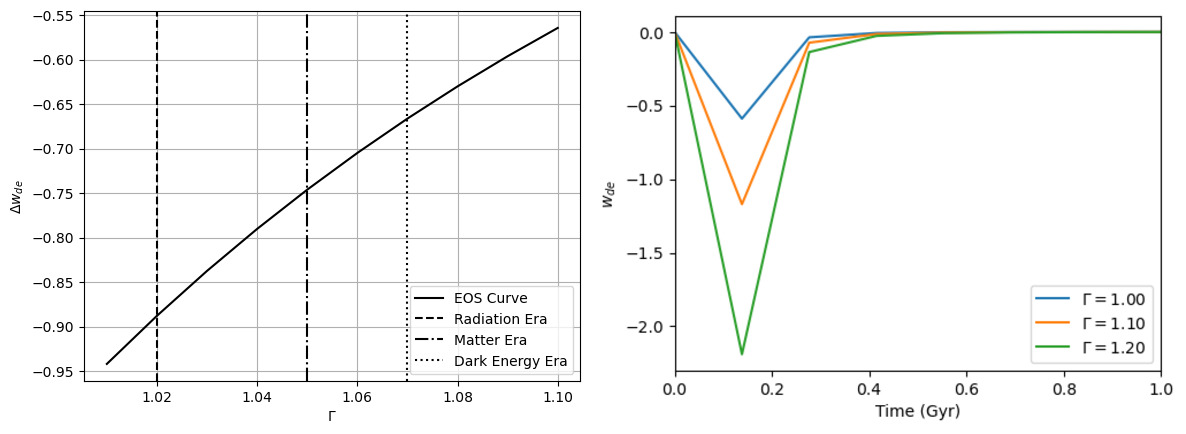}
    \subcaption{Shows the relationship between $\Gamma$ and $w_{de}$ in GCG model: The first plot on the left shows the constrained value of $\Gamma$ in the range $1.00<\Gamma\leq 1.10$ for the early universe. The second plot on the right shows a drastic variation in EoS at around 0.2 Gyr visible for almost all values of $\Gamma$. This is expected for the formation of stars and galaxies in the early universe peaked at 3.5 Gyr after the Big Bang \citep{Salcido2018}. Note that a high value of $\Gamma$ may cause the GCG to exhibit phantom behavior, which might indicate modified gravity at early times.}
   
  \end{subfigure}
  \begin{subfigure}[b]{0.47\textwidth}
    \includegraphics[width=\textwidth]{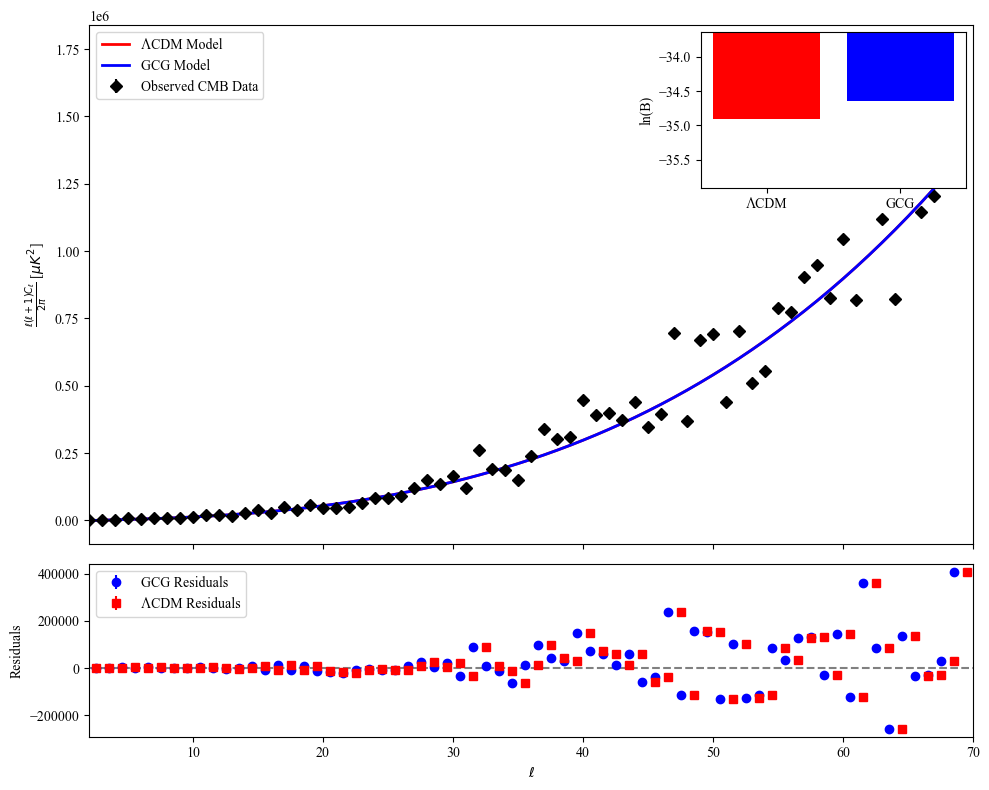}
    \subcaption{Comparing the observed CMB TT power spectrum data with an interacting GCG model and a $\Lambda$-CDM model for $\ell<$ 70. The inset of the bar chart shows the natural logarithm of the Bayesian evidence (ln(B)) for each model, indicating their fit to the data. }

  \end{subfigure}
  \hfill
  \begin{subfigure}[b]{0.5\textwidth}
    \includegraphics[width=\textwidth]{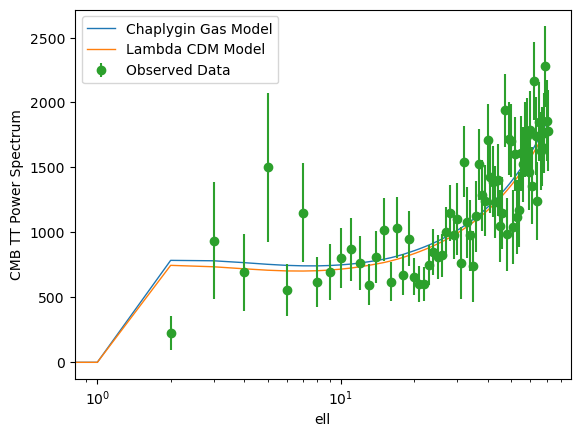}
    \subcaption{The plot compares the interacting GCG model and the $\Lambda$-CDM model against the CMB-TT power spectrum. The GCG with an interaction term could also fit the data for the $\ell < 70$ range, with chi-square values of $69.06$ and $69.93$, respectively.}
   
  \end{subfigure}
  \caption{}
  \label{fig:gcg}
\end{figure} To calculate the energy density of radiation, matter, and dark energy at different epochs, we consider the variation of the scale factor $a$ and the conformal time $d\tau$ w.r.t physical time $t$ and density parameter $\Omega$. 

The behavior of $\Gamma$ changes in GCG, especially at lower multipoles of CMB radiation (refer to Fig. \ref{fig:gcg}(b)(c)). It varies for different epochs. Suppose if the Universe starts with $\Gamma_{\text{chap}} <1$. After inflation, $\Gamma_{\text{chap}}>1$ is significant for early matter fluctuations in CMB. However, any modifications exceeding $|B|<10^{-6}$ have been ruled out in \citep{fabris2011ruling}, and it is now seen as necessary to consider scenarios where only the contribution from the dark energy term dominates (with negligible values for dark matter terms).

The Universe can be approximated as a blackbody radiation field during the radiation epoch. Therefore, $\rho_m \sim 0$ and $\rho_{r} \sim \rho_{\text{critical}}$. The scale factor evolves as $a \propto t^{1/2}$, when the radius of interaction $R\xrightarrow{}R_{radiation}$ and the conformal time are given by $d\tau = dt/a(t) = dt/t^{1/2}$. For a constrained $\Gamma_{\text{chap}} \sim 0.45$, we get the relative density as $$\frac{\Omega_{r}}{\rho_{\text{de}}(t)} \Rightarrow \frac{1}{\rho_{\text{critical}}} =\frac{1}{ \sqrt{\frac{1}{\kappa}R_{radiation}^2\frac{dt}{t^{3/2}}\Gamma_{\text{chap}}^6 + \frac{B}{t}}} \approx 1.28 \times 10^{-26}\ \text{kg m}^{-3}$$ Which is the change in dark energy density during the radiation-dominated era.

During the matter-dominated epoch, the Universe can be approximated as an ensemble of non-relativistic particles. Therefore, $\rho_{m} \sim \rho_{\text{critical}}$ and $\rho_{r} \sim 0$. The scale factor evolves as $a \propto t^{2/3}$, when $R\xrightarrow{}R_{matter}$ and the conformal time is given by $d\tau = dt/a(t) = dt/t^{2/3}$. For a constrained $\Gamma_{\text{chap}} \sim 0.35$, we have $$ \frac{\Omega_{m}}{\rho_{\text{de}}(t)} \Rightarrow \frac{1}{\rho_{\text{critical}}}=\frac{1}{ \sqrt{\frac{1}{\kappa}R_{matter}^2\frac{dt}{t^{4/3}}\Gamma_{\text{chap}}^6 + \frac{B}{t^{4/3}}}} \approx 6.75 \times 10^{-27}\ \text{kg m}^{-3}$$ This shows how the density of dark energy evolves during the matter-dominated era.

At late times, the Universe is dominated by dark energy, so we can assume that $\rho_m =\rho_{r}\sim 0$ and $\rho_{\text{de}} \sim \rho_{\text{critical}}$. During this epoch, the scale factor evolves as $a \propto e^{H_0 t}$, where $H_0$ is the current value of the Hubble constant. We can calculate the dark energy density from the conformal time given by $d\tau = dt/a(t)$ for a constrained $\Gamma_{\text{chap}} \sim 0.01$ as $$\frac{\Omega_{\text{de}}}{\rho_{\text{de}}(t)} \Rightarrow \frac{1}{\rho_{\text{critical}}} =\frac{1}{ \sqrt{\frac{1}{\kappa}R_{\text{de}}^2\frac{dt}{e^{H_0t}}\Gamma_{\text{chap}}^6 + \frac{B}{e^{6H_0t}}}} \approx 2.84 \times 10^{-27}\ \text{kg m}^{-3} $$ Which gives the evolution of dark energy density during the dark energy-dominated era.

The above three scenarios suggest that dark energy density changes with time and depends on the dominant component of the Universe at a given epoch. The relative density was higher during the radiation-dominated era compared to the matter-dominated era, and it reaches its lowest point when dark energy becomes dominant. Changes in dark energy density correspond to the evolution of the scale factor, conformal time, and $\Gamma$. The value of $\Gamma$ is found to be high, ranging from $1.01$ to $1.02$. By constraining this parameter (refer to \ref{fig:gcg}(a)), it is possible to achieve a dark energy density comparable to the matter density without needing to fine-tune the initial conditions.
\subsection{K-essence model}
The K-essence model, originally introduced by Armendariz et al. \citep{armendariz2001essentials} to explain cosmic acceleration, offers an alternative to the standard model by avoiding fine-tuning cosmic parameters and anthropic arguments. In the study of dark energy, the Lagrangian density for the K-essence field is commonly written as a function of the scalar field and its kinetic term. This is similar to the pressure \(p_{de}\), with the Lagrangian density given by \(L = K(X) - V(\phi) = p_{de}\) where \(K(X)\) is the kinetic energy function and \(V(\phi)\) is the potential energy of \(\phi\). However, K-essence can be defined as any scalar field with non-canonical kinetic terms, as these terms can directly relate to the total energy density of the Universe. 

In the general scenario, the equation of motion is given by\citep{armendariz2000dynamical}\citep{scherrer2004purely}:
\begin{equation}
K_{,X}(\Box\phi + V_{,\phi}) = 0
\end{equation}
where $K_{,X}$ is the derivative of the kinetic function ($X= 1/2 (\nabla \phi)^2$) and $\Box\phi = g^{\mu\nu}\nabla_{\mu}\nabla_{\nu}\phi$. 
To express $\rho_{de}$ in terms of the scalar field $\phi$ in flat FLRW, we can take:  
\begin{equation}
\rho_{de} = \frac{1}{2} \Dot{\phi}^2 + V(\phi)
\end{equation} And by using the equation of motion for the K-essence field, we obtain the evolution of the energy density. Further, by taking the time derivative:
\begin{equation}
\dot{\rho}_{de} = \dot{\phi} \ddot{\phi} + V_{,\phi} \dot{\phi}
\end{equation}The equation of motion for the scalar field \(\phi\) in a flat (FLRW) Universe is given by:
\begin{equation}
\ddot{\phi} + 3H \dot{\phi} + V_{,\phi} = 0 
\end{equation} where one can substitute \(\ddot{\phi}\) from the above equation into equation (49):
\begin{equation}
    \ddot{\phi} = -3H \dot{\phi} - V_{,\phi}
\end{equation} To arrive at:
\begin{equation}
\dot{\rho}_{de} = \dot{\phi} (-3H \dot{\phi} - V_{,\phi}) + V_{,\phi} \dot{\phi} = -3H \dot{\phi}^2
\end{equation}\newline\vspace{2mm}
Dividing equation (49) by equation (48), we obtain an equation:
\begin{equation}
\frac{\dot{\rho}_{de}}{\rho_{de}} = \frac{-3H \dot{\phi}^2}{\frac{1}{2} \dot{\phi}^2 + V(\phi)}
\end{equation} We can incorporate the interaction rate parameter \(\Gamma_K\) into the model by modifying the dark energy density evolution equation, assuming an interaction rate \(\Gamma_K\).
\begin{equation}
    \dot{\rho}_{de} = -3H \dot{\phi}^2 + \Gamma_K \rho_{de}
\end{equation}where $\Gamma_{K} \sim \frac{\dot{\phi}}{\sqrt{2}}$ is analogous to the K-essence sound speed, which follows normalization over time evolution of scalar. 

The equation for the time evolution of the energy density is shown in \citep{adams2002fast}. We prefer a transition that depends on \( K \) from \( X \) to \(\phi\), which shows a shift from using kinetic terms to directly modeling the scalar field dynamics in the K-essence framework.

Since $\rho_{de}$ decreases more slowly than the matter density as the Universe expands, it is necessary to express the rate of interaction $\Gamma_K$ in terms of the K-essence field $\phi$ and its time derivative $\dot{\phi}$. Thus, we can relate $\Gamma$ from other alternative models to the K-essence field by the following equation:
\begin{equation}
\Gamma = \frac{\dot{\phi}}{\sqrt{2}} + \gamma \sqrt{\rho_{m}}
\end{equation}
where $\gamma$ is a constant that characterizes the strength of the interaction, and $\rho_{m}$ is the density of matter (including dark matter).

The K-essence model has a fundamental property: it has a negative pressure term once the Universe attains an equilibrium between matter and radiation \citep{armendariz2001essentials}. It overtakes the matter density term to induce cosmic acceleration near the present epoch, potentially affecting the expansion rate. Following this from equation (47), we can compare the expression for $\rho_{de}$ from GCG as:
\begin{equation}
\rho_{de} = \frac{1}{\kappa} R^2(i) \left(\frac{\dot{\phi}}{\sqrt{2}} + \gamma \sqrt{\rho_{m}}\right)^6 d\tau
\end{equation}
which relates the density of dark energy to the K-essence field and the density of matter. Taking the derivative of equation (56) gives the time evolution of $\rho_{de}$, assuming that $R(i)$ and $\gamma$ are constants with respect to time, and $\dot{\rho}_{m}$ is the time derivative of the matter density $\rho_{m}$. This gives a form:
\begin{equation}
\frac{d}{dt}(\rho_{de}) = \frac{1}{\kappa} R^2(i) 6\left(\frac{\dot{\phi}}{\sqrt{2}}\right)^5\left[\ddot{\phi} + \frac{5}{2}\gamma\frac{\dot{\phi}}{\sqrt{\rho_{m}}} \frac{d}{dt}\sqrt{\rho_{m}} + \dots\right]d\tau
\end{equation} where we are interested in $\frac{d}{dt}\sqrt{\rho_{m}}$ that dominates not too long after matter-dominated era. By neglecting higher-order terms and by rewriting the equation terms of $\dot{\phi}$, $\ddot{\phi}$, and $\rho_{dm}$, the change in the dark energy density becomes:
\begin{equation}
\frac{\dot{\rho_{de}}}{\rho_{de}} = 6\left(\frac{\ddot{\phi}}{\dot{\phi}} + \frac{\gamma}{2}\frac{\dot{\rho}_{m}}{\rho_{m}}\right)\frac{1}{\frac{\dot{\phi}}{\sqrt{2}} + \gamma \sqrt{\rho_{m}}}
\end{equation}

It is important to note that the dark energy is negligible during recombination. This can be approximated from: 
\begin{equation}
\begin{aligned}
    \frac{\dot{\rho}_m}{\rho_m} &= \frac{2}{3 t_{rec}} \\
   \ddot{\phi} &\sim 0
\end{aligned}
\end{equation}
 where $\gamma$ , $K(\phi)$ roughly remain a constant. When the scale factor evolves as $a \propto e^{H_0 t}$, we get the fractional change in dark energy density during the matter-dominated era:
\begin{equation}
\frac{\dot{\rho_{de}}}{\rho_{de}} = -3(1+w_{de})H   
\end{equation} and,
\begin{equation}
    w_{de} = -1 - \frac{1}{3} \ (\frac{\dot{\phi}}{\sqrt{2}} + \gamma \sqrt{\rho_{m}} \ )
\end{equation}
This simplified EoS involves scalar fields and interactions between dark energy and matter at roughly the current epoch. The above equation can also be expressed in terms of $\Gamma$ once we substitute equation (55) in equation (61), which gives $$w_{de} = -1 - \left(\frac{1}{3}\right)\Gamma $$ This shows that both GCG and K-essence are equivalent under given conditions to agree with the value of $\Gamma$ unless they describe an effective EoS. 

\begin{figure}[htbp]
    \centering
        \includegraphics[width=0.8\textwidth]{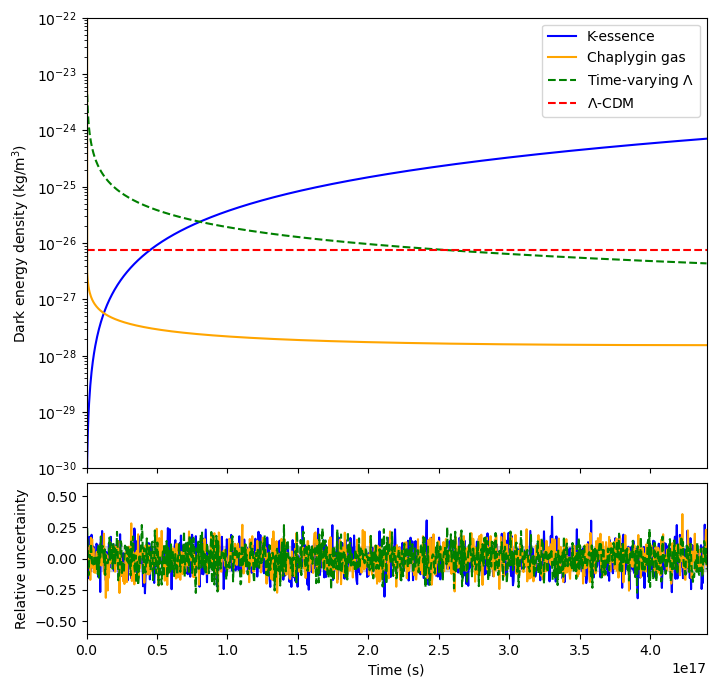}
        \caption {\small Comparison of dark energy densities constrained from different interaction parameters: The K-essence model exhibits evolving estimates for ($\Gamma_K$), starting at ($0.1$) with a ($\pm 10\%$) uncertainty and later narrowing to ($\pm 1\%$). The Chaplygin gas model shows similar behavior, with ($\Gamma_{chap}$) initially estimated at ($1.02$) with a ($\pm 20\%$) uncertainty and later constrained to ($\pm 5\%$). The time-varying cosmological constant model maintains a constant ($\Gamma_{\Lambda}(t) = 0.01$) with a low ($\pm 5\%$) uncertainty across all cosmic epochs. The relatively high uncertainty for Chaplygin gas early in the universe may arise from initial conditions and lack of strong constraints on ($\Gamma_{chap})$.}
        \label{fig:baysien}
    \end{figure}

Otherwise, Fig. \ref{fig:baysien} compares the constrained value of $\Gamma_{K} \sim 0.1$, if $w_{de}<-1$, as is the case for phantom dark energy models, then $1+w_{de}<0$, indicating that the dark energy density increases with time. Conversely, if $w_{de}>-1$, as is the case for quintessence dark energy models, then $1+w_{de}>0$, and the dark energy density decreases with time.
    
\section{K-essence Interacting with GCG}
The Chaplygin gas model has gained attention as a potential candidate for a unifying dark matter and dark energy\citep{kamenshchik2001alternative, avelino2008phys1, avelino2008phys2, avelino2014phys, beca2007mon, liao2013observational, zhang2006new}. This has led to claims that the Chaplygin gas may describe dark matter and energy through a special Lagrangian. However, contradictory results have been found by Sandvik et al. \citep{sandvik2004end}, challenging this claim for unification. Additionally, for a unified model, it is crucial that the equation of state resembles that of the GCG and depends on the scale factor. To demonstrate a unified framework emerging from scalar fields, we consider a non-canonical scalar interacting with GCG, similar to \citep{mishra2021unifying, sadeghi2014interacting}.

We begin by squaring equations (45) and (54), derived earlier, to model this interaction:
\begin{equation}
\rho_{de}^2 = \frac{1}{\kappa}R^2\Gamma_{\text{chap}}^6 d\tau + \frac{B}{a^{6}}
\end{equation} \vspace{2mm}
\begin{equation}
\Dot{\rho_{de}}^2 = 9H^2 \Dot{\phi}^4 - 6H\Dot{\phi}^2 \Gamma_K \rho_{de} + \Gamma_K^2 \rho_{de}^2   
\end{equation}
where we expect a scenario in which GCG, coupled with a scalar field, modifies its rate of evolution dynamically in response to change in background EoS. Both components contribute to the effective energy density and interact linearly. Consequently, we can substitute equation (62) into equation (63) to obtain:
\begin{equation}
    \Dot{\rho_{de}}^2 = 9H^2 \Dot{\phi}^4- 6H\Dot{\phi}^2 \ \Gamma_K  \sqrt{ \frac{1}{\kappa}R^2\Gamma_{{\text{chap}}}^6 d\tau + \frac{B}{a^{6}} }  + \Gamma_K^2 \left( \frac{1}{\kappa}R^2\Gamma_{\text{chap}}^6 d\tau + \frac{B}{a^{6}}\right)  
\end{equation} which is a fourth-order differential equation. The equation can be simplified by introducing an effective interaction parameter $\Gamma_{\text{eff}}$ bound to the range $0 <\Gamma_{\text{eff}} <1$ for constrained $\Gamma_{\text{chap}}$ and $\Gamma_{K}$.

Here, $\Gamma_{\text{eff}}$ resembles the interaction rate term $Q$, where K-essence interacts with the energy density of GCG, combining contributions from both GCG and K-essence:
\begin{equation}
    \Gamma_{\text{eff}} = \Gamma_{K}  \sqrt{ \frac{1}{\kappa}R^2\Gamma_{{\text{chap}}}^6 d\tau + \frac{B}{a^{6}} } 
\end{equation}Or equivalently:
\begin{equation}
      \Gamma_{\text{eff}} = \Gamma_{K}  \sqrt{ A + \frac{B}{a^{6}} } 
\end{equation} Equation (64) further simplifies to:
    \begin{equation}
    \dot{\rho}_{de}^2 = 9H^2 \dot{\phi}^4 - 6H \dot{\phi}^2 \Gamma_{\text{eff}} + \Gamma_K^2 \Gamma_{\text{eff}}^2
\end{equation}From the previous section, we know that $\Gamma_K \sim \frac{\Dot{\phi}}{\sqrt{2}}$, which gives a form: 
\begin{equation}
    \dot{\rho}_{de}^2 = 9H^2 \dot{\phi}^4 - 6H \dot{\phi}^2 \Gamma_{\text{eff}} + \frac{1}{2}\dot{\phi}^2 \Gamma_{\text{eff}}^2
\end{equation} The above equation describes the evolution of dark energy density where the K-essence field velocity (\(\dot{\phi}\)) is modulated by an effective interaction term (\(\Gamma_{\text{eff}}\)). While equation (68) may appear to exhibit features resembling asymptotic behavior under specific conditions, it is not a strictly asymptotic equation in the general sense.

K-essence models often exhibit attractor behavior, where the field naturally evolves towards a solution mimicking a cosmological constant or displaying specific scaling properties \citep{mishra2021unifying, chongchitnan2009cosmological}. Building upon this, we consider a Lagrangian of the form $F(X)-V(\phi)$, with a focus on the kinetic term:
\begin{equation}
\mathcal{L} =  f_0 e^{\lambda \phi} X^n 
\end{equation} and $V(\phi)$ of form 
\begin{equation*}
     V(\phi) = \frac{f_0}{2^n} e^{\lambda \phi} \dot{\phi}^{2n}
\end{equation*} where the K-essence field dynamics is determined by $n$ that gives the power-law behavior to the kinetic term ($X = \frac{1}{2}\dot{\phi}^2$) with $f_0$ as a constant, and \( \lambda \) denotes the coupling strength between the scalar field and kinetic term X. 

\begin{figure}[t]
\centering
    \begin{subfigure}[t]{0.65\textwidth}
        \centering
        \includegraphics[width=\textwidth]{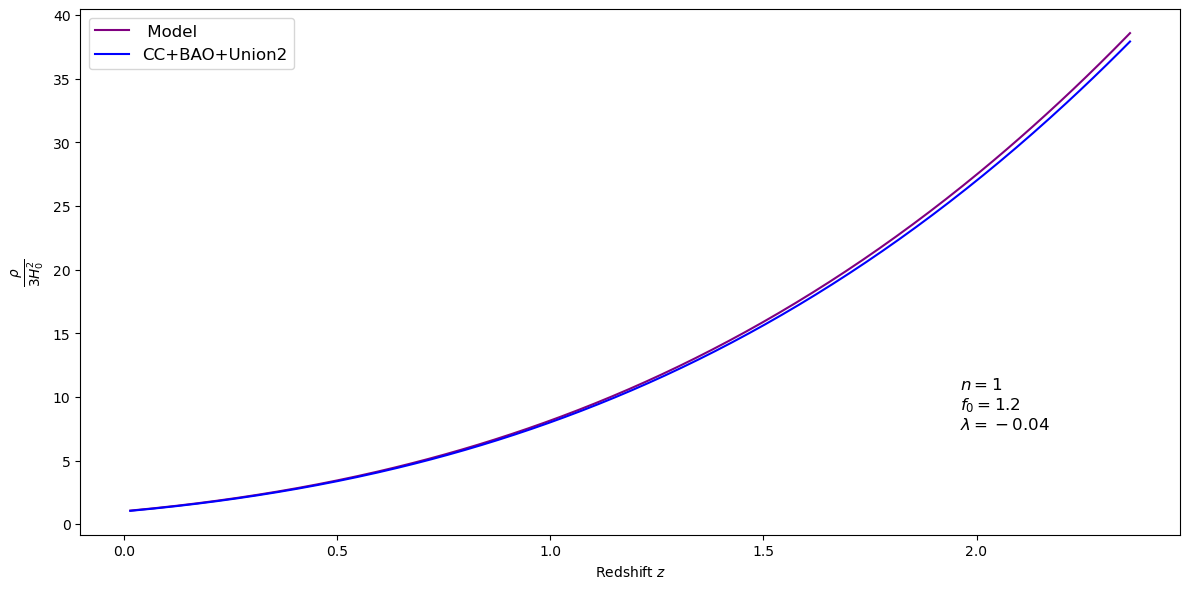} 
        \subcaption{Scalar field density and observed matter density vs redshift.}
        
    \end{subfigure}
    \hfill
    \begin{subfigure}[b]{0.65\textwidth}
        \centering
        \includegraphics[width=\textwidth]{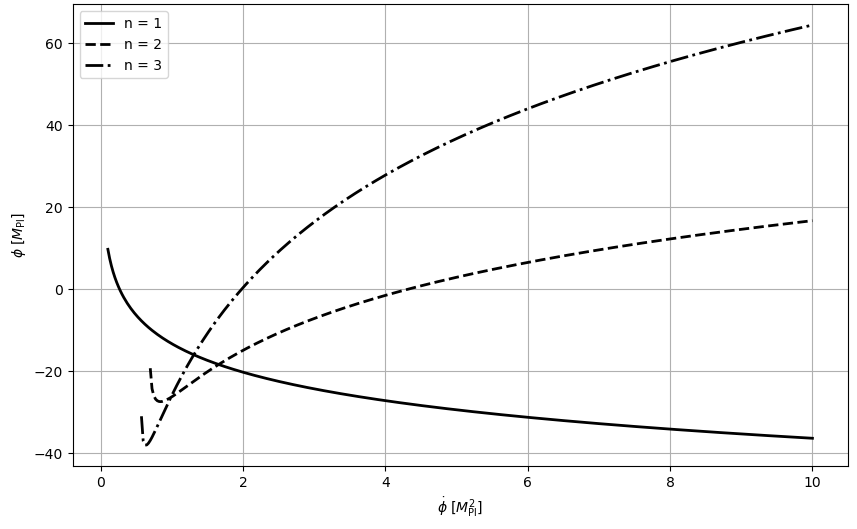} 
        \subcaption{Scalar field dynamics for varying $n$.}
        
    \end{subfigure}
    \caption{}
    \label{fig:curves}
\end{figure}

By obtaining the equations of motion from this Lagrangian and applying the tracker solutions, we can establish the relationship between $\dot{\phi}$ and $H$, which is common in phenomenological models \citep{linton2022momentum}. We start by writing the expressions for energy density \(\rho_{\text{K}}\) and pressure \(P_{\text{K}}\) for the K-essence field:
\begin{equation}
    \rho_{\text{K}} = 2X \frac{\partial \mathcal{L}}{\partial X} - \mathcal{L} = (2n-1)f_0 e^{\lambda \phi} X^n 
\end{equation}
\begin{equation}
    P_{\text{K}} = \mathcal{L} =  f_0 e^{\lambda \phi} X^n
\end{equation} Therefore, the continuity equation becomes:
\begin{equation}
    \dot{\rho}_{\text{K}} + 3H\left(1 + \frac{  f_0 e^{\lambda \phi} X^n}{(2n-1)f_0 e^{\lambda \phi} X^n }\right)\rho_{\text{K}} = -\Gamma_{\text{eff}} f_0 e^{\lambda \phi} n \left( \frac{1}{2} \right)^n \dot{\phi}^{2n}
\end{equation}which can be compared with the standard form:
\begin{equation}
    \dot{\rho}_{\text{de}} + 3H\left(1 + \frac{P_{\text{de}}}{\rho_{\text{de}}}\right)\rho_{\text{de}} \simeq -
    \Gamma_{\text{eff}}\dot{\phi}\frac{\partial \mathcal{L}}{\partial \dot{\phi}}
\end{equation} The kernel $Q$ is found to be:
\begin{equation}
    Q \equiv \Gamma_{\text{eff}}\dot{\phi}\frac{\partial \mathcal{L}}{\partial \dot{\phi}}
\end{equation}where $Q$ is directly proportional to $\Gamma$, with the positive sign indicating flow from dark energy to dark matter. This is valid only for a positive $Q$. If $Q<0$, the dark matter component decays into dark energy, as the equation of motion doesn't hold for $\Gamma$ in equation (33), and ${\phi}$ shifts from negative to positive (for small $n$ and high redshift \citep{vom2017does}) as shown in Figure \ref{fig:curves}(b). Additionally, at the microscopic level, the evolution of dark energy and dark matter is controlled by vastly different scales, and quantum mechanical effects prevent significant interactions between heavy dark matter and light dark energy \citep{d2016quantum}.

Equation (72) can also be expressed in the following form:
\begin{equation}
\dot{\rho}_{\text{K}} + 3H(1 + \frac{1}{2n-1})\rho_{\text{K}} = -\Gamma_{\text{eff}}(2n)\rho_{\text{de}}^{1/2}
\end{equation} where $w_{de} = \frac{1}{2n-1}$ with the speed of sound becoming exceedingly small for large values of $n$ \citep{scherrer2004purely}. The fluid Lagrangian can be used as an attractor. For that, we need to find the initial conditions for $\Gamma_{\text{eff}}$ when the fluid behaves like dust-like matter ( \( \rho_{K} \propto a^{-3}\)). 

During the matter-dominated era, the fluid should satisfy the following conditions:
\begin{equation}
\dot{\rho}_{\text{K}}  = -\Gamma_{\text{eff}}(2n)\rho_{\text{de}}^{1/2}
\end{equation}
where the tracker condition is satisfied when:
\begin{equation}
\Gamma_{\text{eff}} = \frac{3H}{2n}
\end{equation}This has implications, especially during the matter-dominated era. At that time, the K-essence field behaves similarly to that of an attractor, with a slight decrease in the dark energy density in response to the background matter density.
Now, we can use the results obtained from equations (77), (74), and (70) to analyze the behavior of equation (67):
\begin{equation}
    \dot{\rho}_{de}^2 = 9H^2 \dot{\phi}^2 \left(\dot{\phi}^2 - \frac{1}{n} + \frac{1}{8n^2}\right)
\end{equation} The continuity equation becomes:
\begin{equation}
    \dot{\rho}_{de} + 3H \dot{\phi} \sqrt{\dot{\phi}^2 - \frac{1}{n} + \frac{1}{8n^2}} =0
\end{equation}During the matter-dominated era, the fluid acts similar to dust-like matter. The tracking condition is achieved when the effective velocity (the term inside the square root) is proportional to the Hubble parameter, $H$. This allows the fluid to track the background matter density and maintain a fixed interaction rate between the dark energy and matter densities:
\begin{equation}
\dot{\phi} \sqrt{\dot{\phi}^2 - \frac{1}{n} + \frac{1}{8n^2}} = \sqrt{\frac{2}{3}} H
\end{equation} 

Therefore, the tracking solution for the fluid will yield the following form.\vspace{2mm}\newline 
For $n=1$:
\begin{equation}
\dot{\phi} \approx 4H \pm \sqrt{\frac{1}{3(8\dot{\phi}^2 - 7)}}
\end{equation}For $n=2$:
\begin{equation}
    \dot{\phi} \approx 8H \pm \sqrt{\frac{1}{3(32\dot{\phi}^2 - 15)}}
\end{equation}For $n=3$:
\begin{equation}
    \dot{\phi} \approx 12H \pm \sqrt{\frac{1}{3(72\dot{\phi}^2 - 23)}}
\end{equation}With the kinetic term dominating for large $n$, the dynamics of the scalar field are primarily driven by its velocity. The specific form of the potential energy becomes less significant. The scaling of the constant term (inside the square root) with $4n$ shows how the kinetic energy of the K-essence field dominates the equation (80).

To approximate the impact of kinetic terms in $\phi$, we can substitute the density component of equation (79) with equation (70):
\begin{equation}`
    \dot{\phi}^2 \left(\dot{\phi}^2 - \frac{1}{n} + \frac{1}{8n^2}\right) = (2n - 1)^2 f_0^2 e^{2\lambda \phi} \left(\frac{1}{2}\right)^{2n} \dot{\phi}^{4n}
\end{equation}The kinetic energy contribution on the left-hand side balances the combined potential energy and interaction term on the right-hand side, determining the overall evolution of the dark energy density. Solving for different values of $n$, the fluid transitions from a regime where canonical kinetic terms dominate to one where non-canonical kinetic terms and potential-like terms ($f_0$) dominate.\vspace{1mm}\newline
Solving for $n=1$, gives:
\begin{equation}
  \phi = \frac{1}{2 \lambda} \left[ \ln \left( \dot{\phi}^2 \right) - \ln \left(7f_0^2 \right) \right]
\end{equation}Solving for $n=2$, gives:
\begin{equation}
    \phi = \frac{1}{2 \lambda} \left[ \ln \left( 32 \dot{\phi}^{2} - 15 \right) - \ln \left(\dot{\phi}^{6} \right) - \ln \left( 9 f_0^2 \right) \right]
\end{equation}Solving for $n=3$, gives:
\begin{equation}
    \phi = \frac{1}{2\lambda} \left[ \ln \left( 8 (72 \dot{\phi}^2 - 23 )\right) - \ln \left(\dot{\phi}^{10}\right) - \ln \left( 225 f_0^2  \right) \right]
\end{equation}

\begin{figure}
    \centering
    \includegraphics[width=0.65\textwidth]{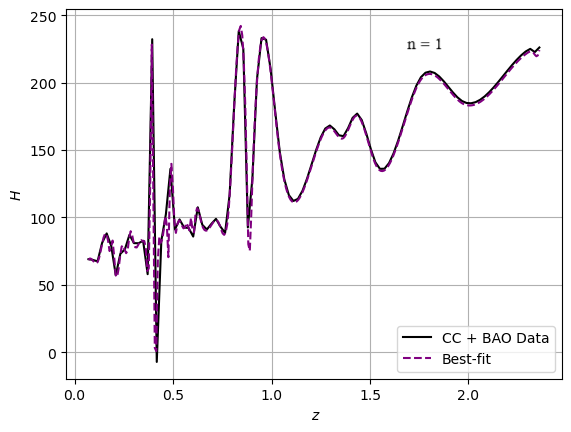}
    \caption{Hubble parameter vs redshift.}
    \label{fig:hvsz}
\end{figure}
For \( n \approx 1 \), the model exhibits characteristics of a balanced regime with significant contributions from dark matter and dark energy shown in Fig \ref{fig:hvsz}. This means the scalar field's density will resemble dark matter at early times (low redshift) and transition towards dark energy at later times (high redshift), similar to the behavior of a unified fluid (GCG) (refer Fig. \ref{fig:curves}). This alleviates the complex dynamical behavior usually seen in conventional models of GCG during late-time acceleration. And for \( n>1 \), the influence of dark energy becomes predominant. 

At late times, as the universe enters the dark energy-dominated epoch, the fluid should behave like a cosmological constant. Here, the term $\Dot{\phi}^2$ become dominant and equation (84) reduces to:
\begin{equation}
\dot{\phi}^4 \approx \sqrt{(2n - 1)^2 f_0^2 e^{2\lambda \phi} \left(\frac{1}{2}\right)^{2n}}
\end{equation} Simplifying the expression inside the square root:
\begin{equation}
    \dot{\phi} \approx \left[(2n - 1) f_0 e^{\lambda \phi} \left(\frac{1}{2}\right)^n\right]^{\frac{1}{2}}
\end{equation}By comparing equation (89) with equation (70), we get a form:
\begin{equation}
   \rho_{\phi} \propto (2n - 1) f_0 e^{\lambda \phi} \left[\left((2n - 1) f_0 e^{\lambda \phi} \left(\frac{1}{2}\right)^n\right)^{\frac{1}{2}}\right]^{2n}
\end{equation} Where we denote $\rho_{\phi}$ to indicate the energy density associated with scalar field. Simplifying further leads to:
\begin{equation}
    \rho_{\phi}  \propto (2n - 1)^{n+1} \left(f_0 e^{\lambda \phi}\right)^{n+1} \left(\frac{1}{2}\right)^{\frac{n}{2}}
\end{equation} This equation describes the energy density of a scalar where the density scales with a power law controlled by the parameter \( n \). 

The EoS $w_{\phi}$ of the scalar take the form: 
\begin{equation}
   w_{\phi} =  \frac{X}{(2n - 1) \dot{\phi}^{2n}}  \approx  \frac{X}{(2n - 1) \left[\left((2n - 1) f_0 e^{\lambda \phi} \left(\frac{1}{2}\right)^n\right)^{\frac{1}{2}}\right]^{2n}}
\end{equation} Which reduces for \( X = \frac{1}{2} \dot{\phi}^2 \), giving:
\begin{equation}
     w_{\phi} \approx  \frac{1}{2} \cdot \frac{1}{(2n - 1)^{2n-1}} \cdot \left[(2n - 1) f_0 e^{\lambda \phi} \left(\frac{1}{2}\right)^n\right]^{2n-1}
\end{equation} As \( \phi \) increases, the term \( e^{\lambda \phi} \) decreases (for negative $\lambda$), which constrains the potential energy contribution. This dynamic transition from the dominance of non-canonical kinetic terms typical in k-essence models to a phase where the potential energy diminishes, resembling Generalized Chaplygin gas (GCG) models.

In a balanced regime ($n=1$) between dark energy and dark matter, the value of $\phi$ is expected to be small at early times. The exponential term becomes \( e^{\lambda \phi} \approx 1 \), resulting in \( w_\phi \) being small, similar to pressureless dark matter (\( w_\phi \approx 0 \)). Any modifications to GCG require $f_0<0$, ensuring a natural transition from dark matter behavior to dark energy behavior as the universe expands. This smooth transition helps explain why dark matter and dark energy densities are comparable today. As the scalar field \( \phi \) evolves and \( e^{\lambda \phi} \) increases, \( w_\phi \) can become increasingly negative, approaching the behavior of dark energy (\( w_\phi \approx -1 \)). The transition depends upon the value of \( f_0 \), similar to how \( B \) in the GCG model determines the transition from dark matter to dark energy. 

\begin{figure}[H]
    \centering
    \includegraphics[width=\textwidth]{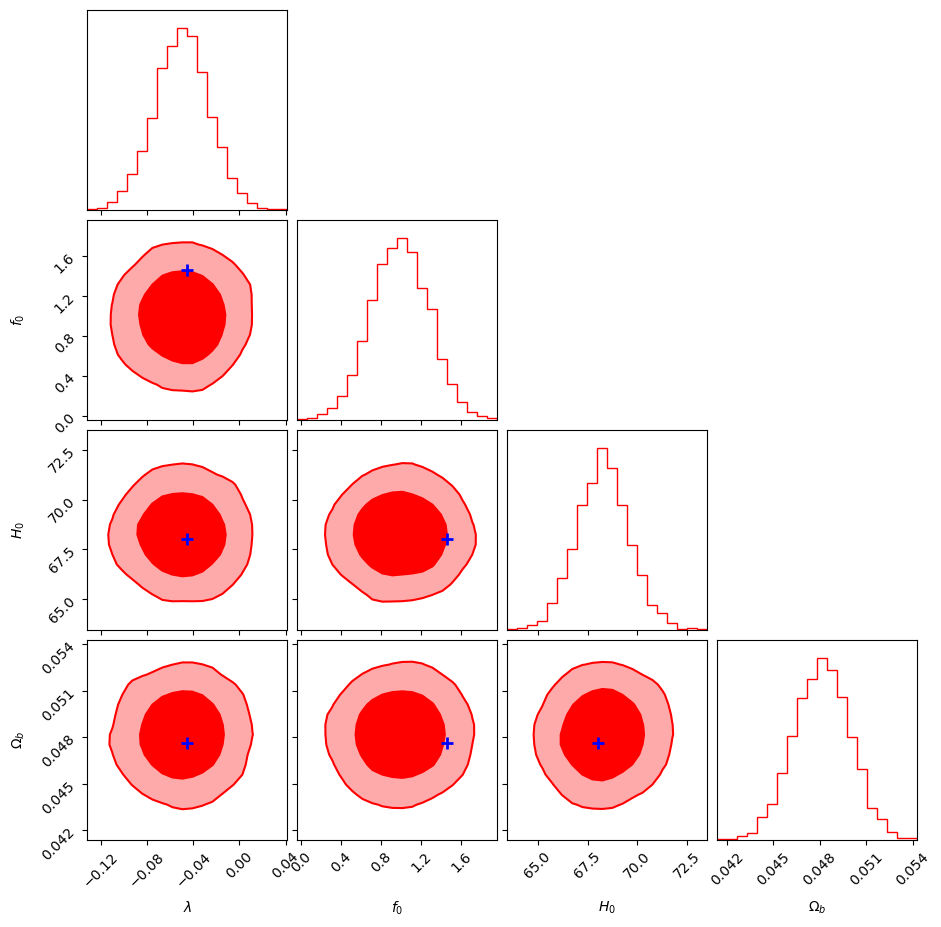}
    \caption{MCMC analysis of CMB data: This plot displays the posterior distributions and correlations of the model parameters (\(\lambda\), \(f_0\), \(H_0\), \(\omega_b\), and \(n=1\)) constrained from the CMB TT-TE power spectra. Using Bayesian analysis, we sampled the parameter space with MCMC, applying priors and computing likelihoods based on the residuals between the theoretical and observed power spectra. The contours represent the 68\% and 95\% confidence intervals, with the maximum likelihood point marked by a blue cross.}
    \label{fig:cmb mcmc}
\end{figure}

\begin{figure}[H]
    \centering
    \resizebox{0.9\textwidth}{!}{ 
        \begin{minipage}{\textwidth}
            \begin{subfigure}[b]{0.25\textwidth}
                \centering
                \includegraphics[width=\textwidth]{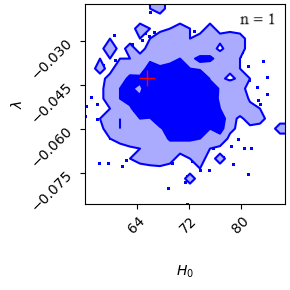}
                \subcaption{}
            \end{subfigure}%
            \begin{subfigure}[b]{0.25\textwidth}
                \centering
                \includegraphics[width=\textwidth]{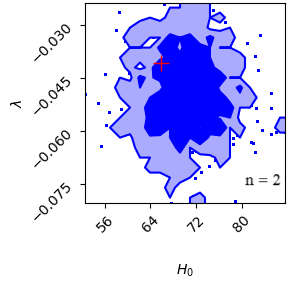}
                \subcaption{}
            \end{subfigure}%
            \begin{subfigure}[b]{0.25\textwidth}
                \centering
                \includegraphics[width=\textwidth]{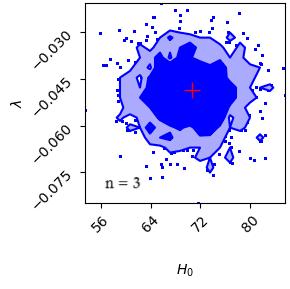}
                \subcaption{}
            \end{subfigure}%
            \vspace{1em}
            \begin{subfigure}[b]{0.25\textwidth}
                \centering
                \includegraphics[width=\textwidth]{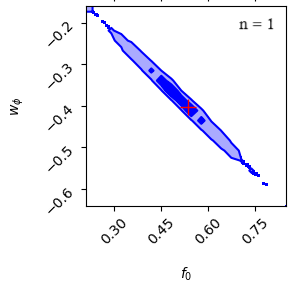}
                \subcaption{}
            \end{subfigure}%
            \begin{subfigure}[b]{0.25\textwidth}
                \centering
                \includegraphics[width=\textwidth]{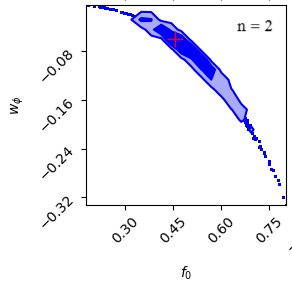}
                \subcaption{}
            \end{subfigure}%
            \begin{subfigure}[b]{0.25\textwidth}
                \centering
                \includegraphics[width=\textwidth]{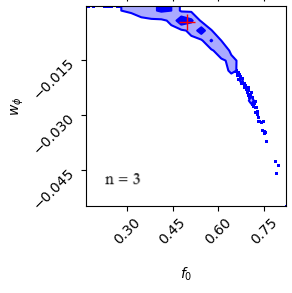}
                \subcaption{}
            \end{subfigure}%
            \vspace{1em}
            \begin{subfigure}[b]{0.25\textwidth}
                \centering
                \includegraphics[width=\textwidth]{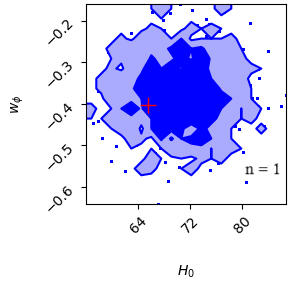}
                \subcaption{}
            \end{subfigure}%
            \begin{subfigure}[b]{0.25\textwidth}
                \centering
                \includegraphics[width=\textwidth]{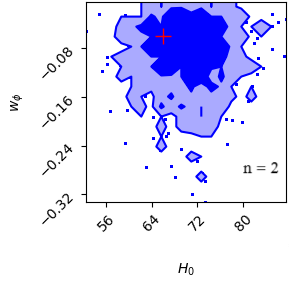}
                \subcaption{}
            \end{subfigure}%
            \begin{subfigure}[b]{0.25\textwidth}
                \centering
                \includegraphics[width=\textwidth]{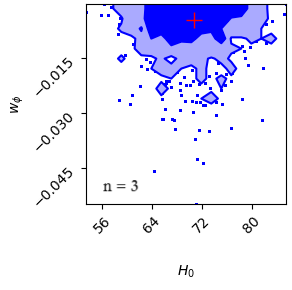}
                \subcaption{}
            \end{subfigure}%
            \vspace{1em}
            \begin{subfigure}[b]{0.25\textwidth}
                \centering
                \includegraphics[width=\textwidth]{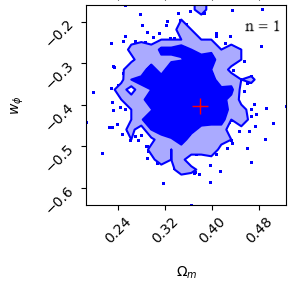}
                \subcaption{}
            \end{subfigure}%
            \begin{subfigure}[b]{0.25\textwidth}
                \centering
                \includegraphics[width=\textwidth]{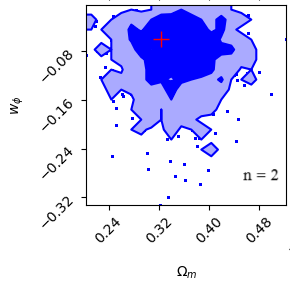}
                \subcaption{}
            \end{subfigure}%
            \begin{subfigure}[b]{0.25\textwidth}
                \centering
                \includegraphics[width=\textwidth]{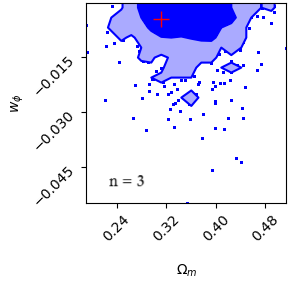}
                \subcaption{}
            \end{subfigure}%
        \end{minipage}    }
    \caption{\small MCMC analysis of Supernovae data: The plot shows the posterior distribution and correlations for model parameters (\(\lambda\), \(f_0\), \(H_0\), \(\Omega_m\)). We employed the Runge-Kutta method, incorporating a likelihood function with the Pantheon + SHOES data. Each subplot illustrates correlations between pairs of parameters, with contours representing the 68\% and 95\% confidence intervals. The maximum likelihood parameters are marked with red crosses.}
    \label{fig:overall}
\end{figure}
\begin{figure}[H]
    \centering
    \includegraphics[width=\textwidth]{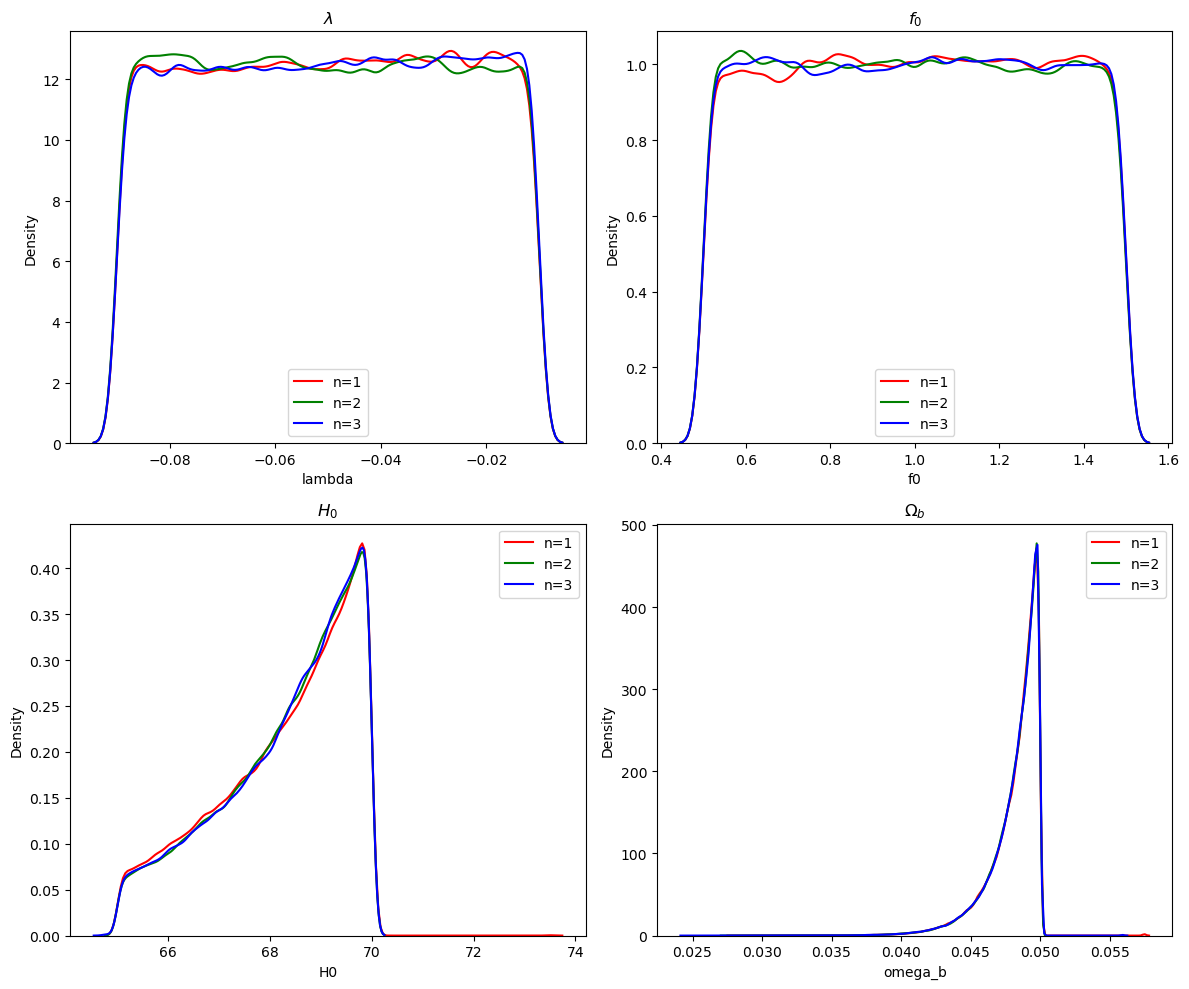}
    \caption{\small This plot shows the posterior probability distributions of four cosmological parameters (\(\lambda\), \(f_0\), \(H_0\), and \(\Omega_b\)) obtained from MCMC sampling in Fig. \ref{fig:cmb mcmc}. Each subplot represents one parameter, with the likelihood curves for different $n$ values that indicate the parameter's distribution constrained from CMB TT-TE power spectrum data.}
    \label{fig:prob}
\end{figure}

\section{Analysis}
Most constraints on IDEs come from large-scale and high-redshift observations, such as cosmic expansion history, CMB, and large-scale structures. These observations can impose strong constraints on IDEs, particularly when the interaction kernel is proportional to dark matter density. In cases where the interaction kernel is proportional to dark energy density, the interaction becomes significant only in the late-time universe, which can have a profound impact on nonlinear structure formation. The ongoing DESI survey \citep{dark_energy_survey, guy2023spectroscopic}, which measures the BAO peak position with high precision, is expected to provide more accurate constraints on the later mentioned. A detailed analysis can be found in \citep{wang2024further}. 

While analyzing different dark energy models, we found that $\Gamma$ for all three models remains positive at late times, although it is less sensitive to dark energy density at early times. In that, the K-essence model is highly sensitive to $\Gamma$ due to its non-canonical kinetic term. The sensitivity of $\Gamma$ has interesting consequences. If the value of $\Gamma$ favors GCG and we measure a high value of $\Gamma_{K} (>0)$ at one epoch, we can rule out the dominance of the K-essence model in that epoch (refer to Fig. \ref{fig:gcg}). Conversely, if $\Gamma$ is measured constant, it would not rule out the K-essence model, as the GCG already has a high value of $\Gamma$ at the same epoch. The time-varying $\Lambda$ maintains a nearly constant value of $\Gamma$ throughout all cosmic epochs. This indicates a relatively weak coupling between dark energy and matter compared to the other two models. However, the CMB constraints of this model may not necessarily rule out the other two models since the value of $\Gamma$ is too small. Including baryonic matter and neutrino interactions (as they can also suppress the matter power spectrum at small scales) can further increase the sensitivity of $\Gamma$.

Observational constraints on the effective interaction parameter (\(\Gamma_{\text{eff}}\)) have also been analyzed. We found that by combining contributions from both K-essence and GCG, $\Gamma$ could fit into the range $0 < \Gamma < 1$, where we have used a specific Lagrangian. The potential of the field $V(\phi) \propto f_0$ is set almost constant while applying CMB constraints. This ensures that the combined fluid behaves like dust-like matter during the matter-dominated era. It is evident from Fig. \ref{fig:curves}(a) that $f_0$ is comparatively higher than we expected from Fig. \ref{fig:overall}, indicating a weak coupling of dark energy with matter at the early stages of the universe. However, upcoming observational facilities, such as DESI, BINGO, SKA, and J-PAS, are expected to provide better constrain for $f_0$. Observations with improved data on redshift-space distortions, cosmic magnification, and galaxy counts may also help to probe the impact of (\(\Gamma_{\text{eff}}\)) out to \(z \leq 3\) and offer a good opportunity to accept or reject certian scalar field Lagrangian \citep{duniya2023cosmological}. 

Further analysis from Fig. \ref{fig:cmb mcmc} and Fig. \ref{fig:prob} reveal that \((\Gamma_{\text{eff}})\) modulates the dynamics of \(\phi\), transitioning from kinetic to potential dominance for \(n < 1\), and modifying the kinetic term behavior for  \(n \geq 1\). This is in conflict with our current understanding of the dark sector. For instance, constraints from SNe and BAO do not directly support $n>1$ \citep{valiviita2010observational}. However, the parameter constraints obtained from MCMC sampling (as shown in Fig. \ref{fig:overall} and Fig. \ref{fig:cmb mcmc}) support these dynamics, with \(\lambda\) determining whether \(\phi\) behaves as a phantom or quintessence field. This demonstrates that the K-essence interacting with GCG can effectively mimic unified dark energy models, acting as dark matter initially and transitioning to dark energy. Particularly at redshift $z \sim 1$, surveys like BINGO and SKA, which are sensitive to this redshift, will be pivotal in probing such transitions of IDEs \citep{costa2019jpas, figueruelo2021jpas, salzano2021jpas}. 

Comparative analysis of model parameters from supernovae and CMB datasets reveals significant deviations in \(w_{\phi}\) and \(f_0\). This might be due to the dynamic nature of \(w_{\phi}\) and \(f_0\), which require further examination. Here, one can assume \(f_0\) acting as constant scaling potential, while \(w_{\phi}\) varies with time. We also found that compared to supernovae data, CMB data better constrains the model parameters, likely due to the additional parameters required to significantly reflect outcomes in local measurements. For the CMB data, the model with balanced regime $n = 1$ gives $H_0 = 68.27 \pm^{1.29}_{1.32}$, which reasonably matches the estimates of the $n = 1$ and $n = 2$ models using supernova data. The $n=3$ model deviates from this agreement as the Eos of the scalar varies significantly within this range,  although the uncertainty calculated for $H_0$ from the supernova data is within 3\%.

The coupling constant $\lambda$ almost agrees with both data sets, but \(f_0\) shows a significant difference of about one magnitude in the CMB data. This is expected since the CMB perturbations can influence the scalar potential, causing it to decrease as the EoS evolves. However, additional data will be required to detect any non-zero coupling between dark energy and dark matter \citep{joseph2023forecast}. On the other hand, the EoS of the scalar field changes significantly, most likely due to decoupling with dark matter for higher values of $n$. Note that we don't put CMB constraints on the EoS but rather assume a quintessence behavior for the scalar field for $n<1$. This adds smoothness to the MCMC estimates since it is challenging to compute the EoS without predicting other CMB parameters.

\section{Discussion}
The main objective of this study is to examine the potential interactions in the dark sector using alternative dark energy models. For this purpose, we introduced the interaction rate $\Gamma$, which is a background-dependent parameter. We then tracked the interaction rate, the entropy of dark energy and matter, and their background densities over conformal time. This allows alternative dark energy models to be extended to incorporate features of other dark energy models while remaining consistent with observational constraints. Within a developed framework, we have specifically examined the time-independent coupling of GCG and K-essence. However, there is no physical reason to assume the coupling will always remain time-independent.

The EoS parameter \( w_\phi \) we studied exhibits diverse behaviors depending on the value of \( n \).  The key is to understand how scalar dark energy models behave and how a balanced regime can result in a controlled EoS that aligns with observations. When \( n = 1 \), \( w_\phi \) can be adjusted to meet observational constraints without deviating much from the currently accepted value \( w \approx -1 \). The equation (93) becomes: 
\begin{equation}
    w_\phi \approx \frac{f_0 e^{\lambda \phi}}{4}
\end{equation} where we set $f_0$ to be negative. By doing so, $\phi$ naturally transitions from a state that behaves like dark matter to one that behaves like dark energy. This offers a hint to solve the cosmic coincidence problem given that $\lambda$ and $w_{\phi}$ are well constrained. When \( n < 1 \), the parameter \( w_\phi \) tends to show behavior similar to quintessence, with \( w_\phi > -1 \), indicating an accelerating expansion. However, the specific behavior and value depend on other model parameters. Conversely, when \( n > 1 \), \( w_\phi \) can exhibit behavior similar to phantom energy, with \( w_\phi < -1 \), resulting in a faster cosmic expansion. 

The interaction between dark energy and dark matter within GCG models often invokes non-physical behaviors, especially at late times. Ensuring a smooth transition to late-time acceleration by avoiding issues such as phantom crossing is a significant challenge. The combined dynamics of the scalar field might mitigate this problem in a balanced regime ($n=1$) where non-canonical and potential terms in unified fluid dominate the late times. Even though this regime uses minimal (K-essence) parameters to capture the dark energy density during late times, the Eos ($w_{\phi}$) demands fine-tuning of $n$, $f_0$, and $\lambda$ to completely avoid divergence. This is often considered a limitation of scalar dark energy models, which we think could be reduced to $n$ and $\lambda$ with further research. However, tuning of these two parameters is now seen as necessary to ensure that this transition is gradual (especially for weak coupling), avoiding any sudden shifts in EoS that could lead to instability or non-physical effects.

When applied to the early universe, the standard GCG model can lead to instabilities, particularly in the matter power spectrum caused by adiabatic pressure perturbation. This can lead to large oscillations or even negative sound speed squared $c_s^2<0$. From our initial consideration of equation (66), one can alleviate this tension by setting $B=0$ ( the dominant GCG term in the early universe) without losing track of matter-dominated era. The effective interaction term becomes:\begin{equation}
    \Gamma_{\text{eff}} \equiv \Gamma_K \sqrt{A}
\end{equation} where $\Gamma_k$ follows normalization over the entire evolution of the scalar field. Note that this adjustment will not change the expected behavior since dark energy in GCG is mainly captured by the $\sqrt{A}$ and matter by $\Gamma_K$ terms. Moreover, GCG would behave like $\Lambda$-CDM in the early universe, with $\Gamma_{\text{eff}}$ primarily dominated by $\Gamma_K$, and GCG maintaining a constant EoS throughout. While $\Gamma_K$ reduces the likelihood of instabilities, it may also result in an incomplete description of the GCG's perturbative dynamics at late times.  

Moreover, our choice of using \(\Gamma\) over $Q$ is motivated by two reasons. Firstly, it can precisely quantify a positive interaction rate (while indicating an accelerated expansion), making it consistent with the Quantum Field Theory framework. A more standard approach involves models that describe dark energy and dark matter as canonical fields with self-interactions. These models typically extend quintessence models by introducing possible interactions with new matter fields. One prototype is a fermion-scalar model that incorporates a Yukawa coupling. For instance, if we consider a process (early universe) where a K-essence particle (dark matter) interacts with a GCG field (dark energy), the scattering amplitude \(\mathcal{M}\) might take the Yukawa form:
\begin{equation}
    \mathcal{M} \sim \int d^4x \, \Gamma_{\text{eff}} \, \bar{\psi} \psi \phi
\end{equation}where \(\bar{\psi}\) and \(\psi\) are fermionic fields representing dark matter particles, and \(\phi\) represents the K-essence scalar field respectively (refer pg. 116-123 of \citep{peskin2018introduction}). The differential cross-section for the scattering process becomes:
\begin{equation}
    \frac{d\sigma}{d\Omega} \propto |\mathcal{M}|^2
\end{equation}
Given the form of \(\Gamma_{\text{eff}}\), the cross-section depends on the dynamical evolution of the K-essence field \(\phi\) and the parameters associated with the GCG. If the kernel \( Q > 0 \) under typical quantum field configurations, it may result in a scenario where the associated interaction probability becomes negative. This can lead to ill-defined results like negative cross-sections. Note that in standard interacting dark energy models, the covariant form given by: 
\begin{equation}
    \nabla_\mu T^{\mu}_{(\text{de})\nu} = -Q u_\nu^{(\text{dm})}/a
\end{equation} where \( u_\nu^{(\text{dm})} \) is the four-velocity of the dark matter fluid, shows Q can take both positive and negative value depending upon the rate of energy density transfer. This form of \( Q \) is phenomenological and not necessarily derived from a fundamental Lagrangian, which is different in our case. 
 
Secondly, if dark energy and dark matter interactions can be mapped onto particle scattering processes, this would allow for the use of Feynman diagrams and other QFT tools to model IDEs at a fundamental level. We used two Lagrangians to model different range of parameters to obtain more reliable constraints on the interaction rate. The parameter \(\Gamma\) can be interpreted here as a scale-dependent interaction rate that potentially reduces the parameter dependency on $B$, despite its criticism in\citep{fabris2011ruling, sandvik2004end}. But scale dependence could also imply that different cosmological epochs might exhibit different interaction behaviors, giving different values for $\Gamma$. This approach is useful (at late times) not only for comparing existing extensions of GCG (such as NGCG) but also for IDEs that may be inherently quantum in nature, pointing toward a framework where both entities are represented by interacting quantum fields. This raises the possibility of defining coupling constants and developing field-theoretic Lagrangians that accommodate non-canonical kinetic terms, such as those found in K-essence models.

Considering that the current main tests of IDEs were limited to using the background dynamics and phenomenological bounds on dark matter particles. Exploration of models where effective interaction terms involving both linear and non-linear perturbations is a promising avenue for future research. Quantum mechanical effects and the vastly different scales may indicate dark energy degrees of freedom might be negligible or can be suppressed at microscopic scales. However, at cosmic scales, the exchange of dark energy could generate an attractive force between a pair of dark matter fermions with an effective potential - a slowly varying interaction term. This is typically realized as much heavier dark matter particles typically in the range $>10^{-3}$eV self-interacting with a relatively lighter dark energy field, as we showed in the prototype.    With self-interaction prototypes (including neutrino interactions), dark energy fermions with mass parameter range ($< H_0$) could become significant even at higher redshift without much suppression as the effective potential remains almost the same through its evolution. With further support from the particle physics community, the future of dark energy indeed lies in these interaction models, which can make notable progress in understanding the dark sector.
 
The IDEs studied in the paper have several limitations. The constraints on IDEs rely on weak and sometimes sparse observational data, which could weaken the robustness of the derived models and their predictions. The GCG interacting with K-essence often requires fine-tuning due to the sensitivity of parameters. This can lead to difficulties in fitting observational data accurately. Additionally, focusing on specific cosmological discrepancies, like the low-$l$ anomaly in the CMB power spectrum, may also limit the generalizability of the findings. Given that we haven't considered structure formation and matter density perturbations in this context. The potential decay of scalar dark energy models often requires careful consideration, which this study may not have fully explored. However, precise constraints on $\Gamma$ can differentiate between models that might otherwise provide similar fits to observational data. If any alternative dark energy models cannot accurately predict the observed data, then we suggest that the corresponding interaction parameter used might differ from the original framework.

\section{Declaration of competing interest}
\begin{enumerate}
    \item Author Contributions: All authors contributed equally to the preparation of the manuscript.
    \item Funding: This research received no external funding.
    \item Data Availability Statement: Not applicable.
    \item Conflicts of Interest: The authors declare no conflict of interest.
\end{enumerate}

\bibliographystyle{ieeetr}  
\bibliography{ref}

\end{document}